\documentclass[12pt,letter]{article}
\usepackage{amsmath,amssymb}
\usepackage{graphicx,color}
\usepackage{hyperref}
\usepackage[bf]{caption}
\renewcommand{\captionfont}{\small\itshape}

\newcommand{\rr}{\mathbb{R}}

\def\theequation{\arabic{section}.\arabic{equation}}

\newcommand{\be}{\begin{equation}}
\newcommand{\ee}{\end{equation}}
\newcommand{\ba}{\begin{aligned}}
\newcommand{\ea}{\end{aligned}}
\newcommand{\ben}{\begin{displaymath}}
\newcommand{\een}{\end{displaymath}}
\newcommand{\bea}{\begin{eqnarray}}
\newcommand{\eea}{\end{eqnarray}}
\newcommand{\bean}{\begin{eqnarray*}}
\newcommand{\eean}{\end{eqnarray*}}

\newcommand{\p}{\partial}

\def\l {\lambda}
\def\th {\theta}
\def\a {\alpha}

\def\b {\beta}

\def\g {\gamma}

\def\e {\epsilon}
\def\s {\sigma}
\def\e {\epsilon}

\def\m{\mu}

\def\k{\kappa}

\long\def\symbolfootnote[#1]#2{\begingroup%
\def\thefootnote{\fnsymbol{footnote}}\footnote[#1]{#2}\endgroup}

\begin{document}

\begin{titlepage}
\vspace{10pt}
\hfill
{HU-EP-09/65}
\vspace{20mm}
\begin{center}

{\LARGE\bf Space-like minimal surfaces in $AdS\times S$}

\vspace{45pt}

{Harald Dorn,$^a~$  Nadav Drukker,$^a~$ George Jorjadze,$^{a,\,b}~$
Chrysostomos Kalousios$^a$
\symbolfootnote[2]{\tt{\{dorn,drukker,jorj,ckalousi\}@physik.hu-berlin.de}}}
\\[15mm]

{\it\ ${}^a$Institut f\"ur Physik der
Humboldt-Universit\"at zu Berlin,}\\
{\it Newtonstra{\ss}e 15, D-12489 Berlin, Germany}\\[4mm]
{\it${}^b$Razmadze Mathematical Institute,}\\
{\it M.Aleksidze 1, 0193, Tbilisi, Georgia}

\vspace{20pt}

\end{center}

\vspace{40pt}

\centerline{{\bf{Abstract}}}
\vspace*{5mm}
\noindent We present a four parameter family of classical string
solutions in $AdS_3\times S^3$, which end along a light-like tetragon
at the boundary
of $AdS_3$ and carry angular momentum along two cycles on the sphere.
The string surfaces are space-like and their projections on $AdS_3$ and
on $S^3$ have constant mean curvature. The construction is based on
the Pohlmeyer reduction of the related sigma model.
After embedding in $AdS_5\times S^5$, we calculate the regularized area
and analyze conserved charges.
Comments on possible relations to scattering amplitudes are
presented. We also sketch time-like versions of our solutions.

\vspace{15pt}
\end{titlepage}
\newpage

\tableofcontents

\section{Introduction}

Minimal surfaces in $AdS_5$ with null polygonal boundary at
conformal infinity have been related to gluon scattering amplitudes
of ${\cal N}=4$ SYM  in a series of papers \cite{Alday:2007hr,
Alday:2008cg, Alday:2009ga, Alday:2009yn} (see also
\cite{Dorn:2009kq, Dorn:2009gq, Jevicki:2009bv, Burrington:2009bh,
Alday:2009dv} for some recent developments beyond $AdS_3\subset AdS_5$).  The
string theory dual to ${\cal N}=4$ SYM lives in $AdS_5\times S^5$.
Therefore, it is a natural extension to look for minimal surfaces in
$AdS_5\times S^5$ with the same null polygonal boundaries for their
projection on $ AdS_5$ but with a non-trivial projection on $S^5$.

One way to approach the problem of constructing classical string
solutions in $AdS\times S$ spaces is through a Pohlmeyer reduction
\cite{pohlmeyer} of the nonlinear string sigma model \cite{devega,Grigoriev:2007bu,Jevicki:2007aa,Jevicki:2008mm,Miramontes:2008wt,Jevicki:2009uz}.  One can view the Pohlmeyer
reduction as a sophisticated gauge choice where we are left with a
model that only involves physical degrees of freedom. This reduced
model inherits integrable structures of the original sigma model.

When including the sphere part of the sigma model, it is no longer
necessary that the projection of the surface on $AdS_5$ be minimal.
Indeed the solutions we find belong to a four-parameter family
of surfaces generalizing both that of \cite{Kruczenski:2002fb,Alday:2007hr} and
\cite{Roiban:2007ju}.
Two of the total four parameters give the mean curvatures of the projections on
$AdS_5$ and on $S^5$. For the $S^5$ projection the corresponding
parameter can also be seen to be related to the  choice of two radii
for a $T^2\subset S^3\subset S^5$. The two remaining  parameters can
be thought of as giving the relative angle and length scale measured
with respect to $AdS_5$ or $S^5$.

String solutions ending on the boundary of $AdS_5$ have a natural
interpretation within the AdS/CFT correspondence as calculating
Wilson loops operators. Furthermore, when the boundary curve is made
of light-like segments, they were interpreted in \cite{Alday:2007hr}
as representing the strong coupling result for gluon scattering
amplitudes in ${\cal N}=4$ SYM.
The space-like solutions presented in this paper
end along tetragons with light-like
edges on the boundary of $AdS_5$. There is a single amplitude for
scattering of four gluons, so the new solutions cannot correspond
directly to such a scattering process. Still, it is always possible
to interpret them in terms of Wilson loops with extra coupling to the
scalar fields. We discuss in particular the example when the extra
scalar insertions are at the cusps. In that case the $AdS_3$ and
$S^3$ parts of the sigma model are aligned.

Examples of strings which end along light-like polygons with more
edges in $AdS_3$ and $AdS_5$ were studied in \cite{Alday:2009yn,
Alday:2009dv}. Generalizations of the solutions presented here along
these lines could have some relevance to the understanding of more
general scattering amplitudes. For more than four gluons there are
helicity dependencies in the amplitude, which so far have not been
realized in string theory. Since our solutions carry extra quantum
numbers, they may be relevant in that regard.

The paper is organized as follows.  In section~2 we formulate the
Pohlmeyer reduction for the construction of minimal surfaces in
$AdS_3\times S^3$ and identify an explicit family of solutions whose
$AdS_3$ projection is space-like and has constant mean curvature. We
also construct a possible $S^3$ projection which makes the surface
minimal, as an object in $AdS_3\times S^3$. Section~3 is devoted to
the study of suitable isometry transformations in $AdS_5$, which
embed our explicit $AdS_3$ surface within the Poincar\'{e} patch of
$AdS_5$. The boundary curve can then realize a generic null tetragon
in Minkowski space.  To discuss a possible relation of our surface to
scattering amplitudes, in section~4 we calculate the regularized
area with respect to the metric  induced from $AdS_5\times S^5$. For
the interpretation of the additional degrees of freedom, arising
from the nontrivial $S^5$ dependence, we study in section~5 some
issues related to conserved currents and related charges.  In
section~6 we discuss an analytic continuation of our surface,
whose boundary is again a light-like tetragon, but with two of the
cusps at time-like separation and the two others space-like
separated.  In section 7 we comment on time-like versions of our
string solutions.  Section~8 contains a summary and some
conclusions.  Several technical issues are presented in appendices.

\setcounter{equation}{0}

\section{The Pohlmeyer reduction in $AdS_3 \times S^3$}

$AdS_3$ is realized as the hyperboloid $\,Y\cdot Y=-1\,$ embedded in
$\rr^{2,2}$. Here $\,Y\equiv(Y_{0'},Y_0,Y_1,Y_2)\,$ denotes
points in $\,\rr^{2,2}\,$ and the scalar product
is given by
\begin{equation}\label{YY=}
Y\cdot Y=Y_1^2+Y_2^2-Y_0^2-Y_{0'}^2~.
\end{equation}
Similarly, $\,S^3\,$ is
realized as $X\cdot X=1$ with the standard metric $\,X\cdot
X=X_m\,X_m$ on $\rr^4$.

In this paper we study space-like surfaces in $AdS_3 \times
S^3$. One can use conformal complex
worldsheet coordinates $z=\frac{1}{2}(\s+i
\tau),~\bar{z}=\frac{1}{2}(\s-i\tau)$.  With the notations
$\p=\p_\s-i \p_\tau$ and $\bar{\p}=\p_\s+i \p_\tau$, the conformal
gauge conditions take the form
\begin{equation}\label{conformal gauge AdSxS}
\partial Y\cdot\partial Y+\partial X\cdot\partial X=
0=\bar\partial Y\cdot\bar\partial Y+\bar\partial X\cdot\bar\partial X~.
\end{equation}
The action functional of the system in this gauge is given by
\begin{equation}\label{S=}
S=\frac{\sqrt\lambda}{4\pi}\int \mbox{d}\sigma \mbox{d}\tau~
\left[\partial Y\cdot\bar\partial Y+\partial X\cdot\bar\partial X
 +\Lambda_1(Y\cdot Y+1)+\Lambda_2(X\cdot X-1)\right]~,
\end{equation}
where $\lambda$ is the coupling constant and $\Lambda_1$,
$\Lambda_2$ are Lagrange multipliers.
The variation of this action leads to the equations
\begin{equation}\label{eq. AdSxS}
\bar\partial\partial Y=\Lambda_1 Y~,~~~~~Y\cdot Y=-1~;~~~~~
\bar\partial\partial X=\Lambda_2 X~,~~~~~X\cdot X=1~,
\end{equation}
which imply the holomorphicity of $\p Y \cdot \p Y$ and $\p X \cdot \p X$
\begin{equation}\label{chirality conditions}
\bar\partial(\partial Y\cdot\partial Y)=0=
\bar\partial(\partial X\cdot\partial X),~~~~~
\partial(\bar\partial Y\cdot\bar\partial Y)=0=
\partial(\bar\partial X\cdot\bar\partial X),
\end{equation}
as well as
\be\label{lambda12}
\Lambda_1=\p Y \cdot \bar{\p} Y, \qquad \Lambda_2=-\p X \cdot \bar{\p} X.
\ee

The 4-cusp solution of \cite{Alday:2007hr} corresponds to the case
$\partial Y\cdot\partial Y=0$ and $\p X=0$.\footnote{Note that for
$\partial X\cdot\partial X=0$ one can have non vanishing $\partial
X$. In particular, the 4-cusp solution of \cite{Alday:2007hr} can be
complemented by a set of instanton solutions on the sphere $S^2$.}
In this paper we consider the case $\partial Y\cdot\partial Y\neq
0$, which implies a non constant spherical part $\partial X\neq 0$.

Using the freedom in the choice of conformal coordinates one can
turn a nonzero holomorphic function $\partial Y\cdot\partial Y$ to
an arbitrary constant, which we take equal to $-1$. This condition
fixes the real $(\sigma,\tau)$ coordinates up to translations
and inversions. In this gauge, the AdS and spherical parts of
the Virasoro constraints are
\begin{equation}\label{gauge fixing}
\partial Y\cdot\partial Y=-1=\bar\partial Y\cdot\bar\partial Y~,~~~~~~~~
\partial X\cdot\partial X=1=\bar\partial X\cdot\bar\partial X~.
\end{equation}

Notice that the $AdS_3$ projection of a space-like surface in
$AdS_3 \times S^3$ may have Euclidean as well as Lorentzian
signature.  Here we study the case where the projection is space-like and clearly the full surface is also space-like.\footnote{The $AdS_3$
projection with Lorentzian signature will be considered in
section 6 (see also \cite{Sakai:2009ut}).}

The Euclidean structure on the $AdS_3$ projection implies
$\partial_\sigma Y\cdot\partial_\sigma Y>0$ and $\partial_\tau
Y\cdot\partial_\tau Y>0.$ These conditions together with
\eqref{gauge fixing} yield $\partial Y\cdot\bar\partial Y>1$.
Similarly, $\partial X \cdot\bar \partial X>1$ and one can use the
parametrizations
\begin{equation}\label{ind-metric AdS}
\partial Y\cdot\bar\partial Y=\cosh\alpha~,~~~~~~~~~~
\partial X\cdot\bar\partial X=\cosh\beta~.
\end{equation}

In this way the description of string  surfaces reduces to a pair of
Euclidean $\sigma$-models given in $AdS_3$ and $S^3$, respectively.
Both systems have a fixed stress tensor given by \eqref{gauge
fixing} and one can apply a Pohlmeyer type reduction \cite
{pohlmeyer}. First we consider the $AdS_3$ projection.

\subsection{The $AdS_3$ projection}

Following the Pohlmeyer reduction, we introduce
a basis in $\rr^{2,2}$ formed by the vectors
$Y,\,\p Y,\, \bar\p Y$ and $N$, where $N$
is the normal vector to the surface
\begin{equation}\label{N dot N}
N\cdot N=-1~,~~~~~~Y\cdot N=0=\p Y\cdot N=\bar\p Y\cdot N~.
\end{equation}
The real vector $N$ is time-like, since the worldsheet
in $AdS_3$ is space-like.

The next step in the reduction is the derivation of the linear
system of differential equations for the basis vectors. These
equations involve the metric tensor $f_{ab}=\p_a Y \cdot \p_b Y$
induced from $AdS_3$ and the coefficients of the second fundamental
form defined by $U_{ab}=-(\p_a\p_b Y\cdot N)$, where
$\p_a=\p_{\xi^a}$ and $\xi^a$ $(a=1,2)$ are worldsheet coordinates.
Introducing the nonzero components of the second fundamental form in
the $(z,\bar z)$ coordinates
\begin{equation}\label{u=}
u=-\p^2 Y\cdot N~,~~~~~~~~\bar u=-\bar\p^2 Y\cdot N~,
\end{equation}
and using \eqref{eq. AdSxS}, \eqref{lambda12}-\eqref{N dot
N}, one finds the following equations
\be\ba\label{system1}
\p^2 Y &= -Y+\frac{\p\a}{\sinh \a}(\cosh\a\,\, \p Y+
\bar{\p}Y)+u N~,\\
\p\bar{\p} Y &= \cosh\a\,\, Y~,\\
\p N&=\frac{u} {\sinh^2\a}\, (\p Y+\cosh\a\,\, \bar{\p}Y)~,
\ea\ee
together with their complex conjugated ones.

The consistency conditions for this linear system give
\be\ba\label{consistency conditions}
\p\bar{\p} \a &=\sinh \a -\frac{u\bar{u}}{\sinh\a},\\
\bar{\p}u &=-\frac{\p\a}{\sinh\a}\, \bar{u}~,
\qquad \p\bar{u}=-\frac{\bar{\p}\a}{\sinh\a}~u.
\ea\ee
These equations have a $(z,\bar z)$-independent solution
\begin{equation}\label{sol}
u=u_0~,~~~~~~~~~\sinh^2 \a =|u_0|^2 ~,
\end{equation}
where $u_0$ is a nonzero complex number. For convenience we
use the parametrization
\begin{equation}\label{u_0=}
u_0=\frac{2}{i}\,\rho\,\sqrt{1+\rho^2}\,\,e^{2i\phi}
~~~~~~~~~(\rho >0)~.
\end{equation}
The induced metric $f_{ab}$ and the
second fundamental form $U_{ab}$ calculated in
the $(\s,\tau)$ coordinates then are
\begin{equation}\label{f=, s=}
f_{ab}=\left(
  \begin{array}{cc}
    \rho^2 & 0 \\
    0 & 1+\rho^2\\
  \end{array}
\right),\quad U_{ab}=
\rho\,\sqrt{1+\rho^2}\,\left(\begin{array}{cc}
    \sin 2\phi & ~~\cos 2\phi \\
    \cos 2\phi & -\sin 2\phi \\
  \end{array}\right)~,
\end{equation}
and the mean curvature, defined as the invariant trace of the
second fundamental form, is
\begin{equation}\label{R=}
H=\frac{1}{2}\,(f^{-1})_{ab}\,U_{ba} = \frac{\sin
2\phi}{2\rho\,\sqrt{1+\rho^2}}~.
\end{equation}
On the other hand the scalar curvature $R$ is zero.

\begin{figure}
\centering
\includegraphics{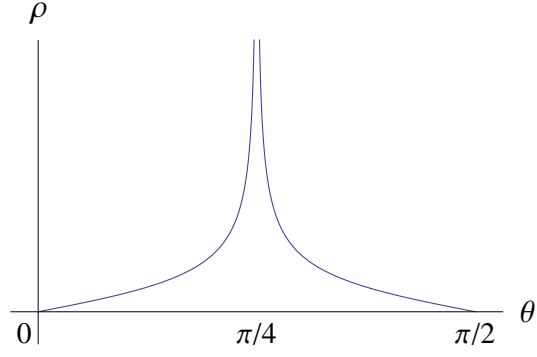}
\caption{{\it The two curves of this plot correspond to
$\tan2\theta=\pm2\rho\sqrt{1+\rho^2}$. The domain bounded by these
curves and the segment $[0,\pi/2]$ on the $\theta$-axis
describes the admissible values of $\rho$ and $\theta$.}}
\label{rhotheta}
\end{figure}

The integration of the linear system \eqref{system1} with $u$ and
$\alpha$ given by \eqref{sol}-\eqref{u_0=} is done in appendix~\ref{AppAdS}.
Up to proper $SO(2,2)$ isometry transformations, the answer is given
by
\begin{equation}\label{AdS sol}
Y_{0'}=\sin \theta\, \cosh \eta~,~~~Y_0=\cos \theta\, \cosh
\xi~,~~~Y_1=\cos \theta \,\sinh \xi~,~~~Y_2=\sin \theta \,\sinh
\eta~,
\end{equation}
where $\theta$ parametrizes the mean curvature of the $AdS_3$ projection
\begin{equation}\label{theta=}
\cot 2\th=H~,~~~~~~~~~~~~~0<\theta<\frac{\pi}{2}~,
\end{equation}
and $(\xi,\,\eta)$ can be treated as worldsheet coordinates obtained
from $(\sigma,\,\tau)$ by the linear transformations
\begin{equation}\label{eta,xi=}
\xi =A\,\s +B\,\tau,\qquad \eta =C\,\s+D\,\tau~.
\end{equation}
The norms of the coefficients $A,$ $B$, $C$, $D$ are uniquely
defined by the parameters $(\rho,\,\theta)$, but there is a freedom
in choice of signs of these coefficients, since the relation between
the angle variables $\phi$ and $\theta$ is not one to one. Here we
present the case $A\geq 0$, $B\geq 0$, $D\geq 0$ and $C\leq 0$,
corresponding to $\phi\in\left(-\pi/ 2,-\pi/ 4\right)$
\be\ba\label{A,B=} A &=
\frac{\rho}{\cos\theta}\,\sqrt{(1+\rho^2)\sin^2
\theta-\rho^2\,\cos^2\theta}~,&
B&=\frac{\sqrt{1+\rho^2}}{\cos\theta}\,\sqrt{(1+\rho^2)\cos^2
\theta-\rho^2\,\sin^2\theta}~,\\
C&=-\frac{\rho}{\sin\theta}\,\sqrt{(1+\rho^2)\cos^2
\theta-\rho^2\,\sin^2\theta}~,&
D&=\frac{\sqrt{1+\rho^2}}{\sin\theta}\sqrt{(1+\rho^2)\sin^2
\theta-\rho^2\,\cos^2\theta}~. \ea\ee It has to be noted that
changing the signs of an even number of these coefficients provides again a solution of the equations of motion \eqref{eq.
AdSxS}, \eqref{lambda12} and the Virasoro constraints
\eqref{gauge fixing} are satisfied as well.

From \eqref{R=} and \eqref{theta=} follow the inequalities
$-1\leq 2\rho\sqrt{1+\rho^2}\,\cot2\theta\leq 1$, which are
equivalent to
\begin{equation}\label{th-rho bound}
\frac{\rho}{\sqrt{1+\rho^2}}\leq
\tan\theta\leq\frac{\sqrt{1+\rho^2}}{\rho}~.
\end{equation}
Due to this bound on possible values of $\theta$ for a
given $\rho$ (see fig.~\ref{rhotheta}), the coefficients of the linear transformations
in \eqref{eta,xi=} are well defined. The boundary points
in fig. \ref{rhotheta} correspond to diagonal or antidiagonal transformations
in \eqref{eta,xi=}.

Note that both fundamental forms \eqref{f=, s=}
are diagonal in the $(\xi,\eta)$ coordinates.
We use these coordinates
in the next sections for calculation of physical quantities.

For $\theta=\pi/4$ the constructed surface
\eqref{AdS sol} coincides with the four cusp solution of
\cite{Kruczenski:2002fb, Alday:2007hr} and, therefore, it is minimal in $AdS_3$.
However, if $\th \neq \pi/4$ the surface \eqref{AdS sol} is
not minimal in $AdS_3$, since its mean curvature \eqref{theta=} is
nonzero.

The surface \eqref{AdS sol} depends on the parameter $\th$ and not
on $\rho$. It can also be written as
\begin{equation}\label{Y_0=}
Y_0=(Y_1^2+\cos^2\th)^{1/2}~,~~~~~~ Y_{0'}=
(Y_2^2+\sin^2\th)^{1/2}.
\end{equation}
This shows that $(Y_1,Y_2)$ can be used as global coordinates
on the surface. Note that \eqref{Y_0=} is one of four parts of the
intersection of the hypersurface $Y_0^2-Y_1^2=\cos^2\th$ and the
$AdS_3$ hyperboloid $Y\cdot Y=-1$.

\begin{figure}
\centering
\includegraphics{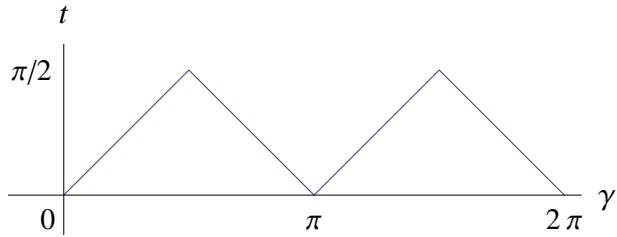}
\caption{{\it The zigzag line here describes the boundary of the
surface \eqref{Y_0=}.}}
\label{tphi}
\end{figure}

\begin{figure}
\centering
\begin{tabular}{c}
\includegraphics{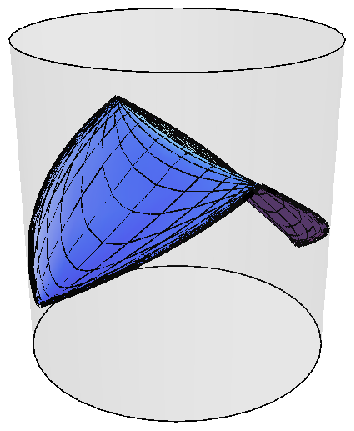}\\$\theta \rightarrow 0$~~~~~
\end{tabular}
\begin{tabular}{c}
\includegraphics{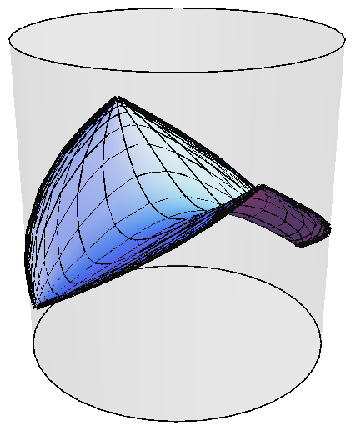}\\$\theta=\pi/ 4$~~~~~
\end{tabular}
\begin{tabular}{c}
\includegraphics{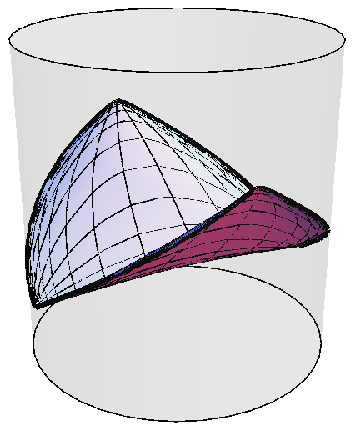}\\$\theta \rightarrow \pi /2$~~~~~
\end{tabular}
\caption{{\it Three different  $AdS_3$ solutions \eqref{AdS sol}.  The plot in the middle corresponds to the solution of \cite{Alday:2007hr}.}}
\label{adsplot}
\end{figure}

Using the global $AdS_3$ coordinates $(q,\,t,\,\gamma)$ defined by
\begin{equation}\label{global coordinates}
Y_{0'}=\cosh q\,\sin t~,\quad Y_0= \cosh q\,\cos t,\quad
Y_1=\sinh q\, \cos\gamma,\quad Y_2= \sinh q\,\sin\gamma
\end{equation}
and taking the limit $q\rightarrow\infty$, we obtain  from \eqref{Y_0=}
that $\cos t=|\cos\gamma|$ and $\sin t=|\sin\gamma|$.
These equations  provide
$\th$-independent light-like segments at the $AdS_3$
boundary plotted in fig.~\ref{tphi}.
Due to the $\theta$ independence, the tetragon formed by these
segments is the same as the boundary of the four cusp solution
in \cite{Alday:2007hr}.
The surfaces \eqref{AdS sol} for different
values of the parameter $\theta$ are presented in
fig.~\ref{adsplot}. One can observe that the surfaces have different
shapes (extrinsic curvature), but the same boundary.

Thus, we conclude that the boundary of the surfaces \eqref{AdS sol}
is $\theta$-independent, though the shape of surfaces inside the
$AdS_3$ depends on the parameter $\theta$.

\subsection{The $S^3$ projection}

Similarly to the $AdS_3$ case, we introduce a basis in $\rr^4$ ($X$,
$\p X$, $\bar\p X,$ $M$) with
\begin{equation}\label{M dot M}
M\cdot M=1~,~~~~~~X\cdot M=0=\p X\cdot M=\bar\p X\cdot M~,
\end{equation}
and find the linear system formed by the equations
\be\ba\label{system2}
\p^2X&=-X+\frac{\p\b}{\sinh\b}(\cosh\b\, \p X-\bar{\p}X)+v M~,\\
\p\bar{\p}X&=-\cosh\b\, X~,\\
\p M&=\frac{v}{\sinh^2\b}(\p X-\cosh\b\,\bar{\p}X)~,
\ea\ee
and their complex conjugated ones.  Here
\be
v=\p^2 X \cdot M, \qquad \bar{v}=\bar{\p}^2 X \cdot M
\ee
are the non-zero components of the second fundamental form.
The consistency conditions for this
linear system lead to the equations
\be\ba\label{consistency condition 2}
\p\bar{\p}\b&=-\sinh\b+\frac{v\bar{v}}{\sinh\b} ~ ,\\
\bar{\p}v&=\frac{\p\b}{\sinh\b}\bar{v}~,\qquad
\p\bar{v}=\frac{\bar{\p}\b}{\sinh\b}v~,
\ea\ee
which similarly to the $AdS_3$ case admit the constant solution
\begin{equation}\label{sol 2}
v=v_0~,~~~~~~~\sinh^2\b=|v_0|^2~.
\end{equation}
In the $(\s,\tau)$ coordinates the corresponding induced metric $h_{ab}=\p_a X \cdot \p_b X$ and the second fundamental form
$V_{ab}=\p_a \p_b X \cdot M$ are
\begin{equation}\label{h=, s=}
h_{ab}=\left(
  \begin{array}{cc}
    1+\rho^{2}_{s} & 0 \\
    0 & \rho^{2}_s \\
  \end{array}
\right),\quad V_{ab}=\rho_s\sqrt{1+\rho_s^2}
\left(\begin{array}{cc}
    \sin2\phi_s & ~~\cos2\phi_s \\
    \cos2\phi_s & -\sin2\phi_s \\
  \end{array}\right),
\end{equation}
where $\rho_s$ and $\phi_s$ parametrize the solution \eqref{sol 2}
\begin{equation}\label{v=}
v_0=\frac{2}{i}\,\rho_s\sqrt{1+\rho_s^2}\,\, e^{2i\phi_s},
\end{equation}
similarly to \eqref{u_0=}.

The mean curvature of the surface in $S^3$  is then given by
\begin{equation}\label{R_s=}
H_s=\frac 1 2 (h^{-1})_{ab}V_{ba}=
-\frac{\sin2\phi_s}{2\rho_s\sqrt{1+\rho_s^2}}~,
\end{equation}
and the scalar curvature $R_s$ is again zero.

The integration procedure presented in appendix~\ref{AppSphere}
leads to the following surface in
$S^3$
\begin{equation}\label{sph sol}
X=(\sin \theta_s \,\cos\eta_s,~\cos \theta_s \,\cos
\xi_s,~\cos \theta_s\, \sin \xi_s,~\sin\theta_s \,\sin
\eta_s)~.
\end{equation}
Here $\theta_s$ parametrizes the mean curvature as in
\eqref{theta=}
\begin{equation}\label{vartheta=}
\cot 2\theta_s=H_s,~~~~~~~~~~~~~~~0<\theta_s<\frac{\pi}{2}~.
\end{equation}
The worldsheet coordinates $\eta_s,\, \xi_s$  are obtained
by the linear transformation
\begin{equation}\label{eta_s=}
\xi_s =A_s\,\s +B_s\,\tau,\qquad \eta_s
=C_s\,\s+D_s\,\tau~,
\end{equation}
with\footnote{Note that the functional dependence of the coefficients \eqref{eta_s=} on $\rho_s$ and $\th_s$ is different from that of the analog coefficients for the AdS part on $\rho$ and $\th$.}
\be\ba\label{sph A,B=} A_s &=\frac{\sqrt{1+\rho_s^2}}{\cos\theta_s}\,
\sqrt{(1+\rho_s^2)\cos^2
\theta_s-\rho_s^2\,\sin^2\theta_s}~,&
B_s&=\frac{\rho_s}{\cos\theta_s}\,\sqrt{(1+\rho_s^2)\sin^2
\theta_s-\rho_s^2\,\cos^2\theta_s}~,\\
C_s&=-\frac{\sqrt{1+\rho_s^2}}{\sin\theta_s}\sqrt{(1+\rho_s^2)\sin^2
\theta_s-\rho_s^2\,\cos^2\theta_s}~,&
D_s&=\frac{\rho_s}{\sin\theta_s}\,\sqrt{(1+\rho_s^2)\cos^2
\theta_s-\rho_s^2\,\sin^2\theta_s}~. \ea\ee
Similarly to the $AdS_3$ case, we have the freedom in choice of signs of the coefficients $A_s, B_s, C_s, D_s$.
The parameters $(\rho_s, \theta_s)$ are
bounded as in \eqref{th-rho bound}, which makes \eqref{sph A,B=}
well defined.

\begin{figure}
\centering
\begin{tabular}{c}
\includegraphics{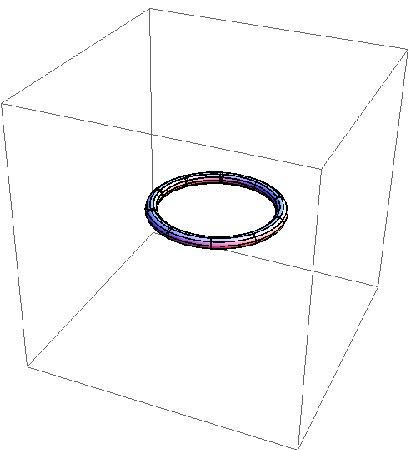}\\$\theta_s \rightarrow 0$~~~~~
\end{tabular}
\begin{tabular}{c}
\includegraphics{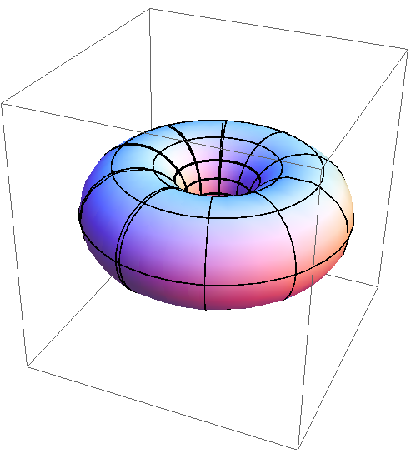}\\$\theta_s=\pi/ 4$~~~~~
\end{tabular}
\begin{tabular}{c}
\includegraphics{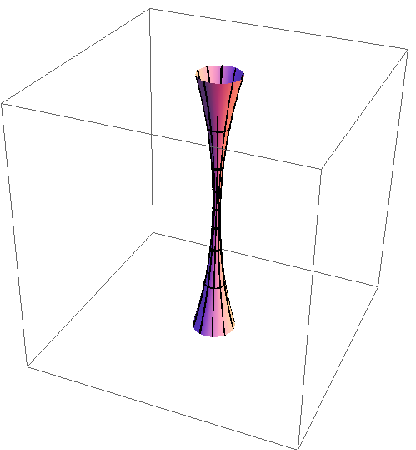}\\$\theta_s \rightarrow \pi /2$~~~~~
\end{tabular}
\caption{{\it Three plots showing a stereographic projection of the
solutions \eqref{sph sol} with different values of the parameter
$\theta_s$. The left plot shows that the case $\theta_s\rightarrow
0$ maps to a circle.  The right plot is the solution for
$\theta_s\rightarrow{\pi/2}$, which is a similar degenerate
torus, but now projected to an infinite line in the stereographic
projection.}}
\label{sphereplot}
\end{figure}

The surface \eqref{sph sol} corresponds to a torus, where $\theta_s$ parametrizes the two radii
of the cycles. Using the stereographic projection of $S^3$
onto $\rr^3$, we present the surfaces \eqref{sph sol} in
fig.~\ref{sphereplot}. The special case $\th_s=0$ or $\th_s=\pi/2$ and $\th=\pi/4$ is related to the solution in \cite{Roiban:2007ju}.

The induced metric tensor in $AdS_3 \times
S^3$, given as a sum of AdS \eqref{f=, s=} and spherical \eqref{h=,
s=} parts, has the form
\begin{equation}\label{induced metric}
g_{ab}=f_{ab}+h_{ab}=(1+\rho^2+\rho_s^2) \left(
  \begin{array}{cc}
    1 & 0 \\
    0 & 1 \\
  \end{array}
\right)~,
\end{equation}
and it is conformal due to our construction. We use  this metric
in section 4 to calculate the worldsheet area.

A last comment concerns the geometric (coordinate invariant) meaning
of the four parameters $\theta ,\theta_s,\rho,\rho_s$. As stated
in (\ref{theta=}) and (\ref{vartheta=}), $\theta$ and $\theta _s$
parametrize the mean curvatures of the projections on $AdS_3$ and on
$S^3$, whereas $\rho $ and $\rho _s$ parametrize the two invariant
quantities $(h^{-1})_{ab}~f_{ba}$ and $(f^{-1})_{ab}~h_{ba}$. Thus,
they control the relative size of lengths on the surface measured
with respect to $AdS_3$ or $S^3$.

\setcounter{equation}{0}

\section{Embedding in $AdS_5$}

We now consider an embedding of the surface \eqref{AdS sol} in
$AdS_5$. The aim is to realize this in such a manner, that the boundary of the surface is inside the boundary of a single Poincar\'{e} patch. To match the generic
configuration of the tetragon boundary, we perform boosts, similarly
to \cite{Alday:2007hr}. However, one has to note that our initial
configuration \eqref{AdS sol} is a rotated version of the surface
used in \cite{Alday:2007hr}. Due to this modification, the relation
of the boost parameters to the Mandelstam variables will be
different.

$AdS_5$ is realized as a hyperboloid in $\rr^{2,4}$
with the coordinates
$Y\equiv(Y_{0'},Y_0,Y_1,Y_2,Y_3,Y_4)$.
We first perform a boost in the $(Y_{0},Y_4)$ plane with a parameter
$b>0$
\begin{equation}\label{boost 1}
Y_{0} \mapsto \sqrt{1+b^2}\,Y_{0}+b\,Y_4~,
~~~~~~~~~~ Y_4 \mapsto
b\,Y_{0}+ \sqrt{1+b^2}\,Y_4~,
\end{equation}
and then a boost in the $(Y_{0'},Y_4)$ plane, which we write in the lightcone form
\begin{eqnarray}\label{boost 2}
Y_{0'}+Y_4 \mapsto \frac{1}{a}\,\left(Y_{0'}+Y_4\right)~,
~~~~~~~~~~ Y_{0'}-Y_4 \mapsto
a\left(Y_{0'}-Y_4\right)~, ~~~~~~(a>0).
\end{eqnarray}
As a result, from \eqref{AdS sol} we
obtain the following surface in $AdS_5$
\begin{equation}\label{boost sol}
Y_0=\sqrt{1+b^2}\,\cos \theta\,\cosh \xi~,
~~~~Y_1=\cos \theta \sinh \xi~,~~~~Y_2 = \sin
\theta \sinh\eta~,~~~~Y_3=0~,
\end{equation}
\begin{equation}\nonumber
Y_{0'}+Y_4=\frac{1}{a}\,(\sin \theta\, \cosh
\eta+b\,\cos\theta\,\cosh\xi)~,~~Y_{0'}-Y_4=a(\sin \theta\, \cosh
\eta-b\,\cos\theta\,\cosh\xi)~.
\end{equation}

Introducing the Poincar\'e coordinates $(r,\,y_\mu),$
$(\mu=0,\,1,\,2,\,3)$, defined as
\begin{eqnarray}\label{Poincare coord}
Y_\mu = \frac{y_\mu}{r}, ~~~~~~Y_{0'}+Y_4=\frac 1 r~,~~~~~
Y_{0'}-Y_4=\frac{r^2 +y_\mu y^\mu}{r}~,
\end{eqnarray}
we find $y_3=0$ (we neglect this coordinate below) and
\be\ba\label{r=}
r&=\frac{a}{\sin\theta\,\cosh\eta+b\,\cos\theta \cosh\xi}~,&
y_0&=\frac{a\sqrt{1+b^2}\,\cos\theta\,\cosh\xi}{\sin\theta\,
\cosh\eta+b\,\cos\theta \cosh\xi}~,\\
y_1&=\frac{a\,\cos\theta\,\sinh\xi}{\sin\theta\,
\cosh\eta+b\,\cos\theta \cosh\xi}~,&
y_2&=\frac{a\,\sin\theta\,\sinh\eta}{\sin\theta\,
\cosh\eta+b\,\cos\theta \cosh\xi}~.
\ea\ee

From the analysis of the previous section follows that
the edges at the boundary are obtained in the limit
$|\xi|\rightarrow \infty$ and $|\eta|\rightarrow \infty$, with finite
$|\xi|-|\eta|.$ Depending on the signs of $\xi$ and $\eta$ there are
four possibilities, which correspond to four edges.
Obviously, $r\rightarrow 0$ in all these cases.

Let us consider the case $\xi>0$, $\eta>0$.
Introducing the parameter $\kappa=\xi-\eta$,
in the limit $\xi\rightarrow \infty$, $\eta\rightarrow \infty$
we obtain
\begin{equation}\label{boundary vectors}
y_0=\frac{a\sqrt{1+b^2}\cos\theta\,e^{\kappa}}{\sin\theta+
b\cos\theta\,e^{\kappa}}~,~~~~~~
y_1=\frac{a\cos\theta\,e^\kappa}{\sin\theta+
b\cos\theta\,e^{\kappa}}~, \qquad y_2=\frac{a\sin\theta}{\sin\theta+
b\,\cos\theta\,e^{\kappa}}~.
\end{equation}
These equations define a light-like segment with a $\theta$-independent
relations
\begin{equation}\label{strightlines}
y_0=\sqrt{1+b^2}\,y_1~, ~~~~~~~b\,y_1+y_2=a~,
\end{equation}
and the coordinate $y_1$ is bounded by $0< y_1 < {a/b}$.
Note that the $\theta$ dependence in \eqref{boundary vectors}
can also be excluded by the  redefinition of the parameter
$e^\kappa \mapsto \tan\theta\,e^\kappa$.
This fact confirms the $\theta$-independence
of the boundary, mentioned in the previous section.
\begin{figure}
\centering
\includegraphics{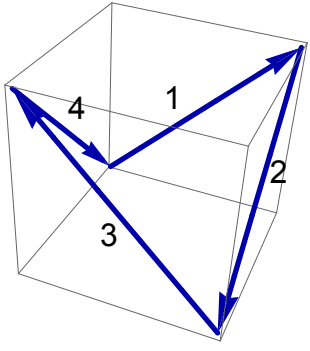}
\caption{{\it The figure here describes the boundary of the surface
\eqref{r=} at $r\rightarrow 0$. The segments with arrows can be interpreted
as vectors in $\rr^{1,2}$, which correspond to \eqref{momenta}.}}
\label{cube}
\end{figure}

Similarly one can take other limits
$\xi\rightarrow \infty$, $~\eta\rightarrow -\infty$, $~\xi+\eta=\kappa$, {\it etc}.

\begin{figure}
\centering
\includegraphics{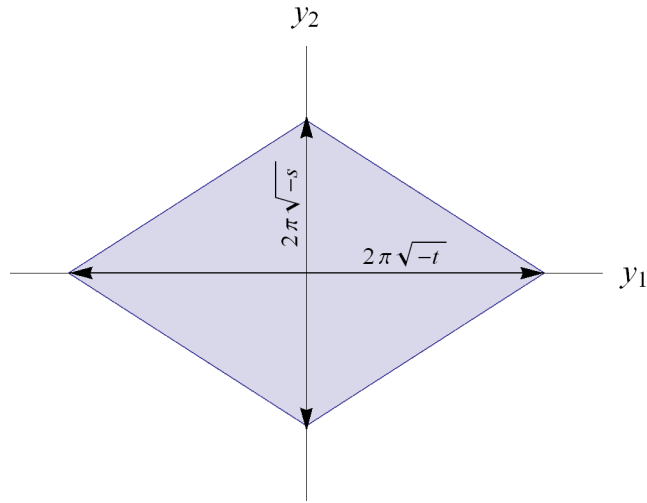}
\caption{{\it This plot shows the projection of fig.~\ref{cube} on the
$(y_1,\,y_2)$ plane.
The diagonals of the rhombus are expressed through the Mandelstam variables.}}
\label{y1y2plane}
\end{figure}

The corresponding segments form the tetragon
given in fig.~\ref{cube}.
The projection of the tetragon
on the  $(y_1,\,y_2)$ plane is the rhombus plotted in fig.~\ref{y1y2plane}.
Its cusps are located on the axes $y_1$ and $y_2$,
and the diagonals are equal to ${2a/b}$ and $2a$, respectively.

The momenta associated with the edges of the tetragon are defined by
$2\pi\,k=\Delta y$,  where $\Delta y$ are the shifts of the $y$ coordinates
along the segments. In this way the four edges define the following
four momenta as in fig. \ref{cube}
\be\ba\label{momenta}
k_1&=\frac{1}{2}\left(\sqrt{-s-t},\,\sqrt{-t},\,-\sqrt{-s}\right),&
k_2&=\frac{1}{2}\left(-\sqrt{-s-t},\,-\sqrt{-t},\,-\sqrt{-s}\right),\\
k_3&=\frac{1}{2}\left(\sqrt{-s-t},\,-\sqrt{-t},\,\sqrt{-s}\right),&
k_4&=\frac{1}{2}\left(-\sqrt{-s-t},\,\sqrt{-t},\,\sqrt{-s}\right)~.
\ea\ee
Here $s$ and $t$ are the Mandelstam variables which are related to the parameters $a,b$ by
\begin{equation}\label{Mandelstam ver}
-s=2k_1\cdot k_2=\frac{a^2}{\pi^2}~,~~~~~~-t=2k_1\cdot k_4=
\frac{a^2}{\pi^2\,b^2}~.
\end{equation}

\setcounter{equation}{0}
\section{Regularized action}

\begin{figure}
\centering
\includegraphics{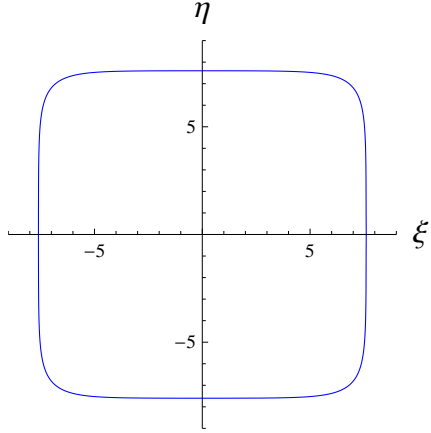}
\caption{{\it The curve is given by \eqref{contour} for the special
case $\epsilon=\epsilon'=0.001$.
It bounds the part of the worldsheet whose area one needs to evaluate in
order to calculate the regularized action  \eqref{reg area}.}}
\label{etaxi}
\end{figure}

Here, we calculate the regularized action for
the string solutions given by \eqref{AdS sol} and \eqref{sph sol}.
We use the scheme proposed in \cite{Alday:2008cg}
and take the worldsheet integration over the domain
$r(\xi,\eta)>r_c$, where the function $r(\xi,\eta)$ is given by
\eqref{r=} and $r_c$ is a small parameter.
According to \eqref{r=} and \eqref{Mandelstam ver},
such a domain is bounded by the contour given as
\begin{equation}\label{contour}
\epsilon\cosh\eta+\epsilon'\cosh\xi=1~,
\end{equation}
where
\begin{equation}\label{epsilon=}
\epsilon= \frac{r_c\,\sin\theta}{\pi\sqrt{-s}}~,~~~~~~~
\epsilon'= \frac{r_c\,\cos\theta}{\pi\sqrt{-t}}~.
\end{equation}
Fig.~\ref{etaxi} shows
the curve \eqref{contour} on the $(\xi,\,\eta)$ plane.\footnote{Note that in the limit $r_c \rightarrow 0$ the edges of the approximate square in fig. \ref{etaxi} correspond to the cusps of the Minkowski space tetragon while the corners are related to the edges of the tetragon.}

With the induced metric tensor
\eqref{induced metric} we get for the regularized action
\begin{equation}\label{reg area}
S_{reg}=\frac{\sqrt\lambda}{2\pi}
(1+\rho^2+\rho_s^2)\int_{r>r_c} \mbox{d}\sigma\,\mbox{d}\tau ~.
\end{equation}
The integral term in \eqref{reg area} is easier to analyze
in the $(\xi,\,\eta)$ coordinates, since the contour which bounds
the domain of integration is symmetric there (see fig.~\ref{etaxi}).
The Jacobian of the transformation to $\xi,~\eta$ coordinates
 defined by \eqref{A,B=} provides
\begin{equation}\label{reg area 1}
S_{reg}=\frac{\sqrt\lambda}{2\pi}\,
\frac{(1+\rho^2+\rho_s^2)\sin2\theta}{\rho\sqrt{1+\rho^2}}\,
I(r_c)~,
\end{equation}
where
\begin{equation}\label{I(r_c)}
I(r_c)=\frac{1}{2}\,\int_{r>r_c}  \mbox{d}\xi\,\mbox{d}\eta~.
\end{equation}
This integral is analyzed in appendix~\ref{AppArea}. Up to terms
vanishing at $r_c\rightarrow 0$ one finds
\begin{equation}\label{I(r_c)=}
I(r_c)= \frac 1 4 \left(\log \frac {r_c^2 \sin^2\th }{-4\pi^2
s}\right)^2 + \frac 1 4 \left(\log \frac {r_c^2\cos^2\th }{-4\pi^2
t}\right)^2- \frac 1 4 \left(\log \frac {s \cot^2\th}{t}\right)^2
-\frac{\pi^2}{3}~.
\end{equation}
For $\th=\pi/4$ this result is in agreement with the corresponding
formula in \cite{Alday:2008cg}.

The expression \eqref{I(r_c)=} can also be written in the form
\begin{equation}\label{I(r_c)=1}
I(r_c)=  \frac 1 4 \left(\log \frac {r_c^2 \cos^2\th }{-4\pi^2
s}\right)^2 + \frac 1 4 \left(\log \frac {r_c^2\sin^2\th }{-4\pi^2
t}\right)^2- \frac 1 4 \left(\log \frac {s}{t}\right)^2 -\frac 1 4
\left(\log \cot^2\th\right)^2 -\frac{\pi^2}{3}~,
\end{equation}
which has the same finite $(s,\,t)$-dependent part as the BDS
formula \cite{Bern:2005iz}. This means that for generic $\th$, $S_{\rm reg}$ in
\eqref{reg area 1}, up to the $\th,\rho,\rho_s$ dependent prefactor,
is given by the result of
\cite{Alday:2008cg}, with a suitable position dependent cutoff.

In order to match with the BDS formula and the strong coupling behavior of the cusp anomalous dimension,
the prefactor in \eqref{reg area 1}
\begin{equation}\label{prefactor}
\frac{(1+\rho^2+\rho_s^2)\sin2\theta}{\rho\sqrt{1+\rho^2}}
\end{equation}
has to be equal to 1.
Due to the inequalities \eqref{th-rho bound}, the minimal value of
$\sin2\theta$ for a given $\rho$ is achieved at the maximal or minimal values
of $\tan 2\theta$ (see fig.~\ref{rhotheta}) and one gets
\begin{equation}
\mbox{min}(\sin2\theta)=
\frac{2\rho\sqrt{1+\rho^2}}{1+2\rho^2}~.
\end{equation}

Hence, the minimal value of the prefactor \eqref{prefactor} is
always greater than 1 and, therefore, there is no finite choice of the
parameters $\rho$, $\rho_s$ and $\theta$ to match with the BDS
formula. One has to note also that the divergent terms in
\eqref{I(r_c)=1} have an asymmetric structure if $\theta\neq\pi /4$.

\setcounter{equation}{0}
\section{Conserved charges}

\subsection{String conserved currents and gluon momenta}

The momenta \eqref{momenta} can be related to the contour integrations of the conserved currents
associated with translations of the Poincar\'e coordinates in
$AdS_5$.

The translations in the Poincar\'e patch $y^\mu\mapsto y^\mu+\epsilon^\mu$
correspond to the isometry transformations in $AdS_5$, which have the
following infinitesimal form
\begin{equation}\label{rotation}
Y^\mu\mapsto Y^\mu +\epsilon^\mu\, Y_+~,~~~~~Y_+\mapsto Y_+~,~~~~~
Y_- \mapsto Y_- +2\epsilon^\nu \,Y_\nu~,
\end{equation}
with $Y_\pm = Y_{0'}\pm Y_4$.
The corresponding Noether currents given by
\begin{equation}\label{current}
J^{\mu}_a=Y_+\,\partial_{a}Y^\mu-
Y^\mu\,\partial_{a} Y_+~,
\end{equation}
define the  1-forms
\begin{equation}\label{1-form}
J^\mu=J_a^\mu\,d\xi^a=Y_+\,dY^\mu-Y^\mu\,dY_+~.
\end{equation}
In  terms of the Poincar\'e coordinates
\eqref{Poincare coord}, these currents become
\begin{equation}\label{1-form=}
J^\mu=\frac{dy^\mu}{r^2}~.
\end{equation}
Their integration over a contour with constant $r$
is trivially performed and gives the displacement of
$y^\m$ along the contour. Such a contour with $r=r_c$ on the
worldsheet \eqref{boost sol} is given by \eqref{contour}.

Let us introduce the $r_c$-dependent charge-like object
\begin{equation}\label{charge}
Q_1^\mu(r_c)=\frac{r^2_c}{2\pi}~\int_{C_1} J^\mu~,
\end{equation}
where the integration contour $C_1$ corresponds to the curve
\eqref{contour} in the first quarter of the $(\xi,\,\eta)$ plane
(see fig.~\ref{etaxi}).
From \eqref{1-form=} follows $Q_1^\mu(r_c)=y^\mu(\xi_c,0)-y^\mu(0,\eta_c)$.
The function $y^\mu(\xi,\eta)$ here is given by \eqref{r=}
and its arguments $(\xi_c,0)$ and $(0,\eta_c)$
are the points where the contour \eqref{contour} intersects
the axes $\xi$ and $\eta$, respectively. Recalling then the
definition of the $k_1$ vector in \eqref{momenta} we find
\begin{equation}\label{Q=k}
\lim_{r_c\rightarrow 0} Q_1^\mu(r_c)=k_1^\mu~.
\end{equation}
The other momenta in \eqref{momenta} are obtained
by the shifts of the contour
$C_1$ to the corresponding quarters in the $(\xi,\,\eta)$ plane.
%%%%%%%%%%%%%%%%%%%%%%%%%%%%%%%%%%%%%%%%%%%%%

\subsection{Charges at the cusps}

Now we would like to study some charges carried by the string, which can be
associated to the cusps of the tetragon. As explicit examples we consider
the charges related to the translation currents for the $AdS_5$ part
\eqref{current} and those related to the rotations in the $(X_1,X_4)$ and the
$(X_2,X_3)$ planes for the $S^3$ part.

For the solution \eqref{boost sol} the translation currents \eqref{current}
take the form
\be\ba\label{trans-currents}
J^0_\xi&=-\frac{\sqrt{1+b^2}}{2a}\,\sin2\theta\,\sinh\xi\,\cosh\eta~,&
J^0_\eta&=\frac{\sqrt{1+b^2}}{2a}\,\sin2\theta\,\cosh\xi\,\sinh\eta~,\\
J^1_\xi&=\frac{1}{2a}\,\sin2\theta\,\cosh\xi\,\cosh\eta+
\frac{b}{a}\,\cos^2\theta~,&
J^1_\eta&=-\frac{1}{2a}\,\sin2\theta\,\sinh\xi\,\sinh\eta~,\\
J^2_\xi&=-\frac{b}{2a}\,\sin2\theta\,\sinh\xi\,\sinh\eta~,&
J^2_\eta&=\frac{b}{2a}\,\sin2\theta\,\cosh\xi\,\cosh\eta+
\frac{1}{a}\,\sin^2\theta~. \ea\ee Here the upper index $\mu $ refers to the
embedding space $(\mu=0,1,2)$, and the lower index to the worldsheet.

The rotational currents for the $S^3$ projection  \eqref{sph sol} have their
simplest form in the $(\xi_s,\,\eta_s)$ coordinates
\begin{equation}\label{s-currents}
J^{14}_{\xi_s}=0~, \quad J^{14}_{\eta_s}=\sin\theta_s~; \quad  \quad
J^{23}_{\xi_s}=\cos\theta_s~, \quad J^{23}_{\eta_s}=0~.
\end{equation}
The worldsheet boundary analysis is most natural in
$(\xi,\,\eta)$ coordinates. To simplify the discussion, we therefore
restrict ourselves to the special case where $(\xi_s,\,\eta_s)$ and
$(\xi,\,\eta)$ have a diagonal transformation matrix. Then the parameters
in \eqref{eta,xi=} and \eqref{eta_s=} have to satisfy
\be\label{A=A_s}
\frac{A}{B}=\frac{A_s}{B_s}\,,\qquad
\frac{C}{D}=\frac{C_s}{D_s}\,.
\ee
These conditions are fulfilled for $B=B_s=C=C_s=0$,\footnote{Besides the reversed situation with minimal value of $\th$ for a fixed $\rho$ {\it etc}, there are no other solutions.} which give the maximal value of $\th$ for a
fixed $\rho$, where $\tan2\theta=-2\rho\sqrt{1+\rho^2}$
(see fig.~\ref{rhotheta}) and minimal value of $\theta_s$ for a fixed $\rho_s$, where $\tan2\theta_s=2\rho_s\sqrt{1+\rho_s^2}$.
This leads to the following relations
between the worldsheet coordinates
\begin{equation}\label{xi=eta}
\xi=\frac{\sigma}{\sqrt{-\cos2\theta}}=
\sqrt{-\frac{\cos2\theta_s}{\cos2\theta}}\,\,\xi_s~,
\quad
\eta=\frac{\tau}{\sqrt{-\cos2\theta}}=
\sqrt{-\frac{\cos2\theta_s}{\cos2\theta}}\,\,\eta_s~.
\end{equation}
The $(\xi,\,\eta)$-components of the currents  \eqref{s-currents}
then read
\begin{equation}\label{s-currents1}
J^{14}_{\xi}=0~,
\quad J^{14}_{\eta}=\sqrt{-\frac{\cos2\theta}{\cos2\theta_s}}\,\,
\sin\theta_s~;  ~~~\quad
J^{23}_{\xi}=
\sqrt{-\frac{\cos2\theta}{\cos2\theta_s}}\,\,\cos\theta_s~,
\quad J^{23}_{\eta}=0~.
\end{equation}
Furthermore, for these special extremal correlations between $\theta$ and
$\rho$ as well as between $\theta_s$ and $\rho_s$, the prefactor in the action
\eqref{prefactor} can be written as
\begin{equation}
\frac{(1+\rho^2+\rho_s^2)\sin2\theta}{\rho\sqrt{1+\rho^2}}
=1-\frac{\cos2\theta}{\cos2\theta_s}~,
\end{equation}
which has a simple form in terms of the charges
we will find below in \eqref{cusp fluxes 1}.
For $\theta\sim {\pi/4}$
and generic $\theta_s$ this is a small deformation away
from unity, which
is the strong coupling limit of the universal scaling function,
or cusp anomalous
dimension. One may hope that it could be related to some generalization of
the cusp anomalous dimension similar to that discussed in
\cite{Freyhult:2007pz,Roiban:2007ju}.

To get an overview of the sources and sinks of the fluxes related
to the worldsheet vector fields  \eqref{trans-currents} and \eqref{s-currents1},
we analyze the corresponding flux lines.

For the rotational currents these
are trivially identified as straight lines parallel to the $\eta$-axis or
parallel  to the $\xi$-axis for $J^{14}$ or for $J^{23}$, respectively.
This means in both cases that all flux originates from one cusp of the tetragon
and flows to the corresponding opposite cusp. There are no other sources
or sinks of flux, neither at other points of the boundary nor in the interior
of the tetragon surface.

\begin{figure}
\centering
\begin{tabular}{c}
\includegraphics{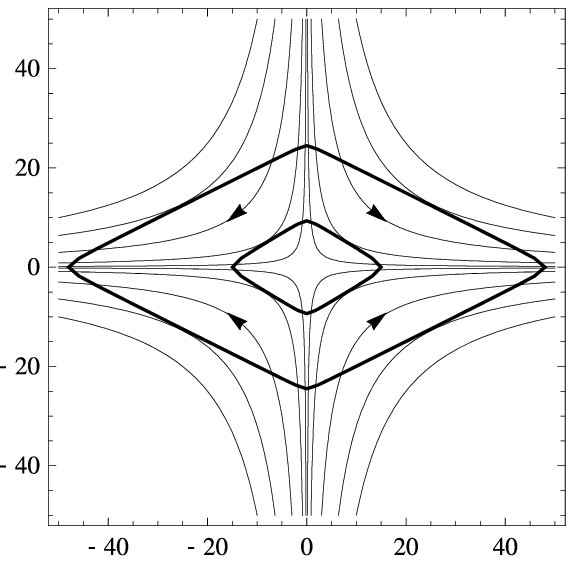}
\end{tabular}
\begin{tabular}{c}
\includegraphics{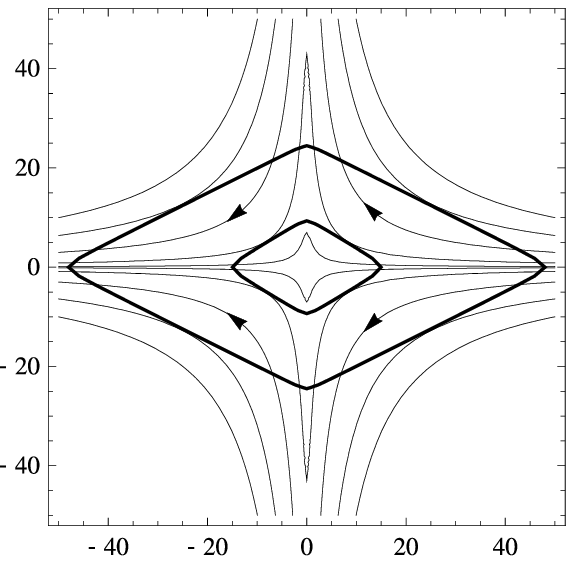}
\end{tabular}
\caption{{\it Regularized boundary (thick curve) in the $(x,y)$ plane for two
values of $(\epsilon,\epsilon')$ with some flux lines for $J^0$ (left) and $J^1$ (right).}}
\label{fluxplot}
\end{figure}

The fluxes of the translational currents require a little bit more effort.
The flux lines for $J^0$ from \eqref{trans-currents} are given by
\begin{equation}\label{J0flux}
\sinh \xi ~\sinh \eta ~=~c~.
\end{equation}
The set of flux lines is parametrized by $-\infty <c<\infty $.
For the two other currents we restrict to large $\xi$ and/or $\eta$. Then
in \eqref{trans-currents} the constant terms can be neglected and the
integration becomes simple again. For $J^1$ and  $J^2$ the flux lines are
given by
\begin{equation}\label{J1flux}
\cosh\xi ~\sinh\eta~=~c~~~\mbox{and}~~~\cosh\eta ~\sinh\xi~=~c~,
\end{equation}
respectively. These fluxes are most efficiently visualized in the plane
spanned by the variables $x=\sinh\xi$ and $y=\sinh\eta$, see fig. \ref{fluxplot}. There the flux lines of
$J^0$ are hyperbolas. The flux lines of $J^1$ behave for large
$x$ like hyperbolas, but cross the $y$-axis at finite values. For
$J^2$ the role of $x$ and $y$ in this picture is interchanged.

From this analysis we see that also the fluxes of the translation currents
have their sources and sinks at the cusps, exclusively. Following standard
terminology, we call these  sources and sinks charges. Then the current
$J^0$ has charges at all four cusps. The sign of the charges alternates,
if one goes around the tetragon. The two other currents have zero
charge at two opposite cusps and charges of different sign at the respective
other pair of opposite cusps.

Obviously, the charges are divergent and need some regularization. Of course,
as usual, regularization is ambigous. However, after treating the area problem
via a regularized boundary according to \eqref{contour}, it is natural to
generate the regularization of the charges also via this regularized
boundary. Therefore, we divide the regularized boundary into charged and
uncharged segments. A negatively charged segment is understood as a connected
piece of  the regularized boundary, on which all flux lines cross from inside
and flow to the same cusp. A positively charged segment is a connected
piece of  the regularized boundary, on which all flux lines cross from outside
and originate from the same cusp. An uncharged segment is a connected
piece of  the regularized boundary, on which all  crossing flux lines
come back at another crossing point.

As the regularization of the charge
sitting at a cusp we define the flux through the charged segment of the
regularized boundary, to which the cusp is connected via its flux lines. For
the flows under discussion this division in segments is exhaustive and a cusp
is connected via its flux lines at most to one charged segment. If it is not
connected to a charged segment, its charge is zero.

The flux through a contour $C$ is given by
\begin{equation}\label{Flux}
F_C=\int_{C} \mbox{d}\xi^a\,\epsilon_{ab}\,\sqrt{g}\,\,g^{bc}\,J_c~=~\int_{C}
\mbox{d}\xi\,\,J_\eta-\mbox{d}\eta\,J_\xi~.
\end{equation}
Note that the rightmost representation is valid since for the subclass of
surfaces under consideration, due to \eqref{xi=eta}, $\xi$ and $\eta $ are
conformal coordinates.

Now we have to identify the points on the regularized boundary, which
define the division into segments. In the $(x,y)$ plane the regularized
boundary \eqref{contour}, \eqref{epsilon=} is given by
\begin{equation}\label{contourxy}
\epsilon ~\sqrt{1+y^2}~+~\epsilon'~\sqrt{1+x^2}~=~1~.
\end{equation}
Asymptotically, for $r_c\rightarrow 0$, this is a rhombus with corners
at the axes at $\pm 1/\epsilon ' $ and $\pm 1/\epsilon$, respectively
\footnote{In contrast to the picture in the $(\xi,\eta )$ plane, the corners
now correspond to the cusps of the AdS tetragon surface.}. It is
shown together with the fluxes for $J^0$ and $J^1$ in fig. \ref{fluxplot}.

For $J^0$
one gets four charged segments. The points defining the division
are determined by the condition, that there the corresponding edge of the
rhombus is tangential to one of the flux line hyperbolas $x~y=c$.
The coordinates of these four points turn out as $(\pm\frac{1}{2\epsilon '},
\pm\frac{1}{2\epsilon})$ (up to terms staying finite for $r_c\rightarrow 0$).

Let us enumerate the cusps in clockwise manner, starting with that
related to the positive $y$-axis. Then $F_1^0$, the regularized charge for
$J^0$ at cusp 1, is given by \eqref{Flux}, where the contour $C$ is the piece
of the regularized boundary in the $(\xi,\eta)$ plane between the points
$(\mbox{arcsinh}\ \frac{1}{2\epsilon '},\mbox{arcsinh}\ \frac{1}{2\epsilon})$
and $(-\mbox{arcsinh}\ \frac{1}{2\epsilon '},\mbox{arcsinh}\
\frac{1}{2\epsilon})$. Due to flux conservation the integration contour
can be replaced by the straight line connecting these two points. Then using
\eqref{trans-currents}, \eqref{Mandelstam ver} and \eqref{epsilon=}  we get
\begin{equation}\label{charge1}
F^0_1~=~\sqrt{-s-t}~\frac{\pi}{2r_c^2}~+~O(1)~.
\end{equation}
Now we define rescaled (renormalized) translation charges via (compare
to \eqref{charge})
\begin{equation}\label{qF}
q^{\mu}_a=\lim_{r_c\rightarrow 0}\frac{r_c^2}{2\pi}~F^{\mu}_a
\end{equation}
and get after repeating the calculation for the other cusps and the currents
$J^1$ and $J^2$
\be\ba\label{cusp charges}
q^0_1&=\frac{1}{4}\ \sqrt{-s-t}~,& q^1_1&=0~,& q^2_1&=\frac{1}{4}\
\sqrt{-s}~,\\
q^0_2&=-\frac{1}{4}\ \sqrt{-s-t}~,& q^1_2&=\frac{1}{4}\ \sqrt{-t}~,&
q^2_2&=0~,\\
q^0_3&=\frac{1}{4}\ \sqrt{-s-t}~,& q^1_3&=0~,& q^2_3&=-\frac{1}{4}\
\sqrt{-s}~,\\
q^0_4&=-\frac{1}{4}\ \sqrt{-s-t}~,& q^1_4&=-\frac{1}{4}\ \sqrt{-t}~,&
q^2_4&=0~.
\ea\ee

For the rotation currents the regularized boundary has only two charged
segments, each segment consists out of two adjacent edges of the rhombus. The
currents are constant, giving regularized charges proportional to
the length of the integration contours in the $(\xi,\eta)$ plane, i.e.
proportional to $\log r_c$. If one would apply the  same rescaling
recipe as for the translation currents, one would get zero renormalized
charges for all cusps.   We prefer to apply a rescaling, which just cancels
the logarithmic divergence and get
\be\ba\label{cusp fluxes 1}
q^{14}_1&=
\sqrt{-\frac{\cos2\theta}{\cos2\theta_s}}~\sin \theta _s ~, &q^{23}_1&=0~,\\
q^{14}_2&=0~, & q^{23}_2&=\sqrt{-\frac{\cos2\theta}{\cos2\theta_s}}~\cos \theta _s~,~~~~~~\\
q^{14}_3&=-\sqrt{-\frac{\cos2\theta}{\cos2\theta_s}}~\sin
\theta _s ~, &q^{23}_3&=0~,\\
q^{14}_4&=0~, &q^{23}_4&=-\sqrt{-\frac{\cos2\theta}{\cos2\theta_s}}~\cos \theta
_s~.
\ea\ee
Note that $q^1$ and $q^2$ are parallel to
$q^{23}$ and $q^{14}$ from (\ref{cusp fluxes 1}) as four-vectors.

Another interesting point is the relation between the charges
\eqref{cusp charges} and the gluon momenta \eqref{momenta} that can be written in contravariant coordinates as
\begin{equation}\label{sp-momenta}
k_n^0~=~ q_{n+1}^0-q_n^0~, \quad \quad
 k_n^i~=~2 \ (q_{n+1}^i-q_n^i)~, \quad (i=1,2)~,\quad (n=1,...,4)
~~(q_5=q_1)~,
\end{equation}
which is quite similar to the result of the previous subsection.
%%%%%%%%%%%%%%%%%%%%%%%%%%

\setcounter{equation}{0}
\section{Analytic continuation of the $AdS_3$ projection}

Here we describe how to get a new space-like
string solution in $AdS_3\times S^3$ with a time-like $AdS_3$
projection from \eqref{AdS sol}.

Let us continue analytically the two parameters $\rho \mapsto -i
\rho$ and $\th \mapsto i \th$ that the $AdS_3$ projection \eqref{AdS
sol} depends on. This procedure leaves the worldsheet coordinates
$\xi$ and $\eta$ real, but makes the components $\,Y_{0'}\,$ and
$\,Y_2\,$  pure imaginary and thus they effectively exchange positions.  Then, the new solution of the system
\eqref{S=} is
\be\label{anal1}
Y_{0'}= \sinh \th \, \sinh\eta, \quad Y_0=\cosh\th \,
\cosh \xi, \quad Y_1=\cosh\th \, \sinh\xi,
\quad Y_2=\sinh\th \, \cosh\eta,
\ee
where
\begin{equation}\label{anal2}
\xi =A\,\s +B\,\tau,\qquad \eta =C\,\s+D\,\tau,
\end{equation}
with
\be\ba\label{anal3}
A &= \frac{\rho}{\cosh\theta}\,\sqrt{(1-\rho^2)\sinh^2
\theta-\rho^2\,\cosh^2\theta}~,&
B&=\frac{\sqrt{1-\rho^2}}{\cosh\theta}\,
\sqrt{(1-\rho^2)\cosh^2
\theta-\rho^2\,\sinh^2\theta}~,\\
C&=\frac{\rho}{\sinh\theta}\,\sqrt{(1-\rho^2)\cosh^2
\theta-\rho^2\,\sinh^2\theta}~,&
D&=\frac{\sqrt{1-\rho^2}}{\sinh\theta}
\sqrt{(1-\rho^2)\sinh^2
\theta-\rho^2\,\cosh^2\theta}~.
\ea\ee
Note that the signs of all coefficients in this  equation are now positive.
However, there is a freedom to change signs of an even number
of the coefficients, like for the solution \eqref{AdS sol}.

The parameters $\rho$ and $\th$ are constrained in such a way that
the square roots in \eqref{anal3} are real, which means
\begin{equation}\label{<rho,theta>}
\rho^2\leq \frac{\sinh^2\theta}{1+2\sinh^2\theta}~.
\end{equation}

\begin{figure}
\centering
\begin{tabular}{c}
\includegraphics{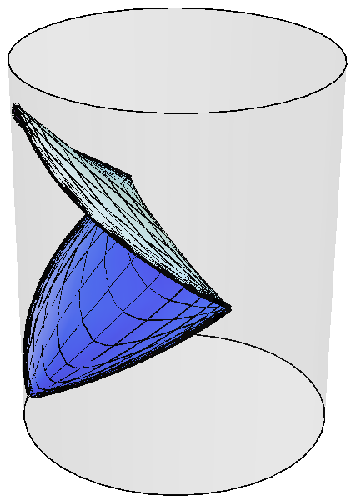}\\$\theta \rightarrow 0$~~~~~
\end{tabular}
\begin{tabular}{c}
\includegraphics{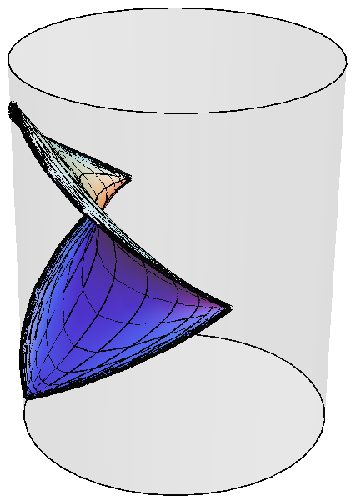}\\$\theta=1$~~~~~
\end{tabular}
\begin{tabular}{c}
\includegraphics{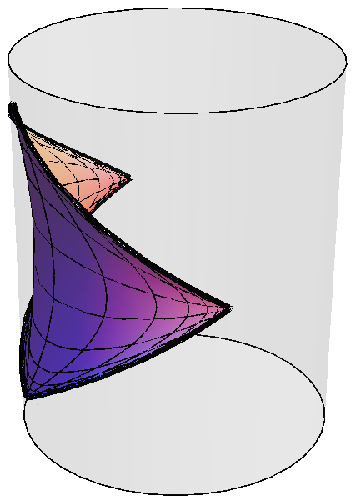}\\$\theta \rightarrow \infty$~~~~~
\end{tabular}
\caption{{\it The $AdS_3$ projection of different solutions \eqref{anal1}, which are
the analytical continuation of the solutions in section 2.
Now there are two cusps
which are time-like separated.}}
\label{adsplot2}
\end{figure}

The induced metric for the $AdS_3$ projection is time-like and in
the $(\s,\tau)$ coordinates it is given by $f_{ab}={\rm
diag}(-\rho^2,1-\rho^2)$. The full surface, including the
contribution from the sphere \eqref{sph sol} renders the total
induced metric Euclidean and conformal, with conformal factor
$1-\rho^2+\rho_s^2$. The new $AdS_3$ projection is visualized in
fig.~\ref{adsplot2} for different values of $\th$.

As opposed to the case studied in section~2, here pairs of consecutive edges
go forward in time, so one pair of the cusps has a time-like separation.

The solution \eqref{anal1} can also be obtained in a systematic way
by the Pohlmeyer reduction. One has to use again complex conformal
coordinates, which for the time-like $AdS_3$ projection yield
$-1\leq\partial Y\cdot \bar\partial Y\leq 1$. Then, one can use the
parametrization $\partial Y\cdot \bar\partial Y=\cos\alpha$, which
modifies the linear equations \eqref{system1} and the consistency
conditions \eqref{consistency conditions}. Note also that the normal
vector $N$ is now space-like.  Finally, a constant solution of the consistency condition, after exponentiation leads to \eqref{anal1}.

\setcounter{equation}{0}
\section{Time-like surfaces in $AdS _3\times S^3$}
The example of the previous section inspires a construction of
time-like worldsheets in $AdS_3\times S^3$.   In this case the
conformal worldsheet coordinates $z=\frac 1 2 (\s+\tau)$ and
$\bar{z}=\frac 1 2 (\s-\tau)$ are real.
\begin{figure}
\centering
\begin{tabular}{c}
\includegraphics{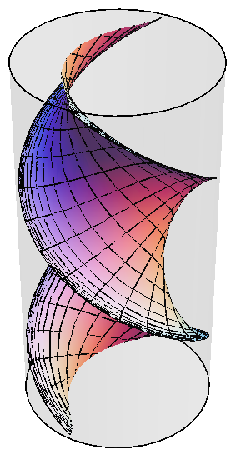}\\(a)
\end{tabular}
\begin{tabular}{c}
\includegraphics{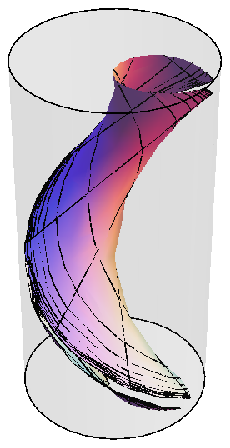}\\(b)
\end{tabular}
\begin{tabular}{c}
\includegraphics{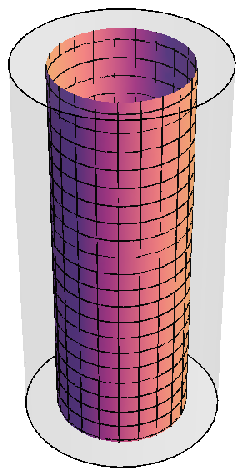}\\(c)
\end{tabular}
\caption{{\it AdS projection of time-like surfaces in $AdS \times S$.}}
\label{timeplot}
\end{figure}
The components of the stress tensor become chiral and they can be turned to a
constant \be\label{virasoro-t}
\p Y \cdot \p Y=-\m^2=\bar{\p} Y
\cdot \bar{\p} Y,\quad \p X \cdot \p X=\m^2=\bar{\p} X \cdot
\bar{\p} X ~.
\ee

Here one has to distinguish infinite open strings and closed strings. In the case of infinite strings the worldsheet is described by a plane and using a rescaling
of worldsheet coordinates one can put $\mu=1$. We present two
solutions for infinite strings. Their structure is defined by
elements of the $o(2,2)$ algebra that one needs to exponentiate in the
integration procedure of the linear system.

The first solution is the hyperbolic case (fig.~\ref{timeplot}a)
\be\ba\label{hyperbolic}
Y_{0'}&= \sinh\th\, \sinh \eta\,
\cos\xi-\cosh\th\, \cosh\eta\, \sin\xi,
& Y_0&=\sinh\th\, \sinh\eta \,\sin\xi+\cosh\th \, \cosh \eta \,\cos \xi,\\
Y_1&=\sinh\th\, \cosh\eta\,\sin\xi+\cosh\th\,\sinh\eta\,\cos\xi, &
Y_2&=\sinh\th\,\cosh\eta\,\cos\xi-\cosh\th\,\sinh\eta\,\sin\xi~,
\ea\ee
and the second one is the parabolic (fig.~\ref{timeplot}b)
\be\ba\label{parabolic}
Y_{0'}&=- \eta\,\cos\xi-\sin\xi,&
 Y_0&=-\eta\,\sin\xi+\cos\xi,\\
Y_1&=\eta\,\cos\xi, &
Y_2&=-\eta\,\sin\xi.
\ea\ee The parameter $\theta$ in \eqref{hyperbolic} is again related to the extrinsic
curvature and the new worldsheet coordinates $(\xi,\,\eta)$ are
given by a linear transformation as in \eqref{eta,xi=}. The equation of
motion and the Virasoro constraints \eqref{virasoro-t} provide three
conditions for the coefficients $A$, $B$, $C$, $D$. Taking into
account the $\theta$-dependence, one gets a two parameter family of
solutions, as in the space-like case.

The closed string solutions satisfy the
periodicity condition $\sigma\sim \sigma+2\pi$. In this case $\mu$
stands as an additional parameter, but the coefficients $A$ and $C$
become integers. The corresponding worldsheets  are defined by the
exponentiation of the time-like elements in the $o(2,2)=sl(2)\oplus sl(2)$ decomposition and lead to the
elliptic solution (fig.~\ref{timeplot}c)
\be\label{periodic} Y_{0'}= \cosh \th \, \sin\eta,
\quad Y_0=\cosh\th \, \cos \eta, \quad Y_1=\sinh\th \, \sin\xi,
\quad Y_2=\sinh\th \, \cos\xi~, \ee where
\begin{equation}\label{k,l}
\xi =k\,\s +\k\,\tau,\qquad \eta =l\,\s+\l\,\tau,
\end{equation}
with integers $k$ and $l$. The equation of motion and the Virasoro
constraints now provide three equations
\be\ba\label{k,l=}
(l^2+\l^2) \cosh^2\th-(k^2+\k^2)\sinh^2\th&=\m^2 \\
k^2-\k^2=l^2-\l^2, \qquad k \k \sinh^2\th&=l\l\cosh^2\th~,
\ea\ee
and
one can use $k$, $l$ and $\mu$ to parameterize the solutions.

The spherical part of the time-like solutions formally coincides with \eqref{sph sol}. However, the conditions for the coefficients $A_s$, $B_s$, $C_s$, $D_s$ now are modified. In the periodic case one has
\begin{equation}\label{}
\xi_s =k_s\,\s +\k_s\,\tau,\qquad \eta_s =l_s\,\s+\l_s\,\tau,
\end{equation}
with integers $k_s,\,l_s$ and one finds equations similar to
\eqref{k,l=}
\be\ba
(k_s^2+\k_s^2) \cos^2\th_s+(l_s^2+\l_s^2)\sin^2\th_s&=\m^2 \\
k_s^2-\k_s^2=l_s^2-\l_s^2, \qquad k_s \k_s
\cos^2\th_s&=-l_s\l_s\sin^2\th_s~. \ea\ee Closed string worldsheets
constructed in this way generalize string solutions in
$AdS_3\times S^1$ and $R^1\times S^3$ discussed in \cite{Hoare:2009rq}.

\setcounter{equation}{0}
\section{Conclusions}

We have constructed a four parameter family of
string solutions
in $AdS_5\times S^5$ whose boundary approaches the light-like
tetragon and are a generalization of the solution of
\cite{Alday:2007hr}. These  minimal surfaces are  space-like and flat.
Their  projections on each of the $AdS_5$ and $S^5$ have constant
mean curvature. As the surface approaches the boundary of $AdS_5$
it wraps a torus inside $S^5$ an infinite number of times. The solutions
therefore satisfy Neumann boundary conditions on $S^5$.

The construction used the Pohlmeyer reduction for $AdS_3\times
  S^3$ and is based on a  constant solution of the
  reduced system. After knowing the solution, it is instructive to describe
it by a direct ad hoc construction as follows.
Up to isometry, the tetragon surface of \cite{Alday:2007hr} can be
characterized as a part of the intersection of the
AdS hyperboloid with the hypersurface $Y_0^2-Y_1^2={1/2}$. The
straightforward deformation to $Y_0^2-Y_1^2=\cos ^2\theta $ with the
explicit parametrization (\ref{AdS sol})
is no longer minimal in $AdS_3$. However, this deformation does not modify
the behavior of the hypersurface at infinity, resulting in the same boundary
behavior.  With the parametrization \eqref{AdS sol} the surface obeys the equation of motion \eqref{eq. AdSxS},\eqref{lambda12}, but not the pure AdS Virasoro constraints.
After a suitable linear transformation to new coordinates ($\sigma$ and
$\tau$ in the main text) one can achieve
$\partial Y \cdot \partial Y=-1$.
It is therefore necessary to include also a nontrivial projection
on $S^5$
satisfying $\partial X \cdot \partial X=1$ such that
the total Virasoro condition is satisfied $\partial Y \cdot \partial Y +
\partial X \cdot \partial X
=0$. The $AdS_5$ part and the $S^5$ are essentially decoupled, their
only correlation is given via the Virasoro constraint \cite{Grigoriev:2007bu}.

Applying suitable isometry transformations in $AdS_5$, the boundary can
fit any null tetragon spanned by four light-like vectors pointing alternating
forward and backward in time. Interpreting these
vectors as momenta of a four-point
scattering amplitude of massless particles in Minkowski space allows us
to characterize the tetragon
by the Mandelstam variables $s$ and $t$. The area of our surface,
regulated by a constant cutoff in the radial coordinate of an $AdS_5$
Poincar\'{e} patch has been
calculated. Up to a prefactor, depending on the parameters $\theta, ~\rho$ and
$\rho_s$, it is given by the same expression as the area for the pure
$AdS_5$ tetragon surface used in the Alday-Maldacena conjecture with
a position dependent cutoff \cite{Alday:2008cg}.%
\footnote{Given that the surface satisfies Neumann boundary conditions
on the $S^5$ coordinates, the classical action should include an extra
Legendre-transformation  boundary term. This term is proportional to the
circumference of the loop times the relevant momenta and has a
logarithmic divergence that is subleading with respect to the bulk
$\log^2$ divergence.}
The prefactor
is generically larger than one. It approaches the value one only in the
limiting case $\theta =\pi/4,~~\rho \rightarrow\infty$. Then
the $AdS_5$ projection is just the tetragon surface of \cite{Alday:2007hr}
and the limit $\rho \rightarrow\infty$ suppresses the weight of the
$S^5$ projection relative to the $AdS_5$ part.

An additional observation was made via the continuation
$\theta\rightarrow i\theta$ and $\rho \rightarrow -i\rho$ connected
with a Wick rotation for one of the time-like and one of the space-like
$\mathbb R^{2,4}$ coordinates. The resulting minimal surface is still
space-like concerning the metric induced from $AdS_5\times S^5$. However,
the induced metric of its $AdS_5$ projection is time-like.
This projection approaches a null tetragon at the boundary.
In the original null tetragon the sides were consecutively running up and down
in time. Now, after this continuation, these null tetragons contain two cusps
with adjacent momenta pointing in the same direction of time. In the language
of scattering amplitudes the surface would correspond directly to the
physical configuration in the $s$- or $t$-channel and not, as before
continuation, to that in the $u$-channel.

One may think that our solutions are subleading contribution
to the usual gluon scattering amplitude. However, as we have shown
in section~5, these solutions carry extra quantum numbers, which are
momenta on the sphere, or $R$-charge in the gauge theory. These surface
therefore describe different Wilson loop observables and if they describe
scattering amplitudes, then these are also somewhat different.
One would expect the dual Wilson loop to have
insertions of the appropriate scalar fields carrying these charges
(similar to \cite{Drukker:2006xg}).
In the particular setup analyzed in section~5, the $AdS_3$ and $S^3$
parts of the solution are chosen to align.
In that case the charges are localized at the cusps and each
corresponds to rotation in a single plane. Therefore, one would expect the
dual Wilson loop to have insertions of the form
$Z^{F_{14}}$, $X^{F_{23}}$, $\bar Z^{F_{14}}$, $\bar X^{F_{23}}$
where $Z$ and $X$ are two complex scalars. In the more
general case these insertions could be along the edges.%
\footnote{Alternatively, it may involve insertions at the cusps
of appropriate (complicated) operators carrying two angular momenta.}
It would be interesting to study these insertions
on the gauge theory side to see if similar structure arises also in
perturbation theory.

The solutions carrying a single charge at each cusp have the additional
nice feature that in each of the quadrants in the
$(\xi,\,\eta)$ plane both the $AdS_3$ and the $S^3$ coordinates have a
uniform behavior, exponential in the $AdS_3$ part and oscillatory in the $S^3$
part.  In the construction of the solutions for more general light-like
polygons \cite{Alday:2009yn, Alday:2009dv}, such
regions on the worldsheet with uniform asymptotics play a crucial
role. It is natural to expect that similar generalization would apply
to our construction.

It is of course tempting to speculate that such solutions may represent
scattering amplitudes other than the MHV ones, which seem to be
captured purely by the $AdS$ solutions. Clearly this should not be the case
of these specific solutions, since for four gluons all amplitudes are
either MHV or they vanish. Still, since some generalization is necessary
to describe non-MHV amplitudes for higher-point functions,
it is natural to look for solutions
similar to the ones presented here which end on more light-like segments
on the boundary.

In a final addendum, we indicated how our procedure for space-like string
  surfaces can be translated to the study of dynamical strings and gave
  examples for related explicit solutions. Besides a complete description
of all solutions in this class, further study should explore how they fit into previously studied cases and in particular into
  the general classification given in \cite{Beisert:2005bm}.

\section*{\large Acknowledgements}
We thank J. Henn, K. Jin, J. Plefka, C. Vergu, S. Wuttke and D.
Young for useful discussions. This work has been supported in part
by Deutsche Forschungsgemeinschaft via SFB 647. G.J. was also
supported by GNSF.

\newpage
\appendix

\setcounter{equation}{0}
\def\theequation{A.\arabic{equation}}
\section{Integration of the linear system in $\rr^{2,2}$}
\label{AppAdS}

In this appendix we construct the string surface
corresponding to the solution \eqref{sol}-\eqref{u_0=}.

We use covariant notations
$(\sigma,\,\tau)$=$(\xi^1,\xi^2)$;  $\,\p_\sigma=\p_1$, $\p_\tau=\p_2,\,$
and introduce the basis related to these real worldsheet
coordinates
\begin{equation}\label{o-n basis}
e_{0'}=N~,~~~~e_0=Y~,~~~~e_1=\frac{\partial_1
Y}{\rho}~,~~~~e_2=\frac{\p_2 Y}{\sqrt{1+\rho^2}}~.
\end{equation}
From \eqref{f=, s=} follow the orthonormality conditions $e_m
\cdot e_n = \mbox{diag}(-1,-1,1,1).$

For a given $Y$, $\partial_1Y$ and $\partial_2Y$,
the normal vector $N$ can be fixed by
\begin{equation}\label{normal=}
e_{0'}\,^{n}=\epsilon^{n}\,_{mlk}\,e_2\,^{m} e_1\,^{l} e_0\,^{k}~,
\end{equation}
where $\epsilon^{n}\,_{mlk}$ is the Levi-Civita tensor,  with
$\epsilon_{0'012}=1$, and upper letters correspond to
the vector indices of the basis elements. Then,
$e_m^{~~n}\in SO_\uparrow(2,2)$, i.e.
\begin{equation}\label{so}
\mbox{det}\,e_m^{~~n}=1~,
\quad e_{0'}\,^{0'}e_{0}\,^{0}-e_{0'}\,^{0}e_{0}\,^{0'}\geq 1~.
\end{equation}
The vector $\partial_t Y$, where $t$ is the global time variable
\eqref{global coordinates}, defines the time direction.
From \eqref{so} follows
$N\cdot\partial_tY\leq -1$, which indicates that $N$
is oriented to the time direction.

Similarly to \eqref{system1}, the basis vectors \eqref{o-n basis}
satisfy the linear equations
\begin{equation}\label{e'=}
\p_1 e =A_1 \,e~, ~~~~~\qquad \p_2{e} =A_2\, e~.
\end{equation}
The matrices $A_1$ and $A_2$ here belong to the $o(2,2)$ algebra and
they have the following block structure
\begin{equation}\label{matrices}
A_1 = \left(
       \begin{array}{cc}
         0 & B_1 \\
         B_1^T &0  \\
       \end{array}
     \right),
~~~~~~~~~
     A_2 = \left(
       \begin{array}{cc}
         0 & B_2 \\
         B_2^T &0  \\
       \end{array}
     \right),
\end{equation}
with $2\times 2$ matrices
\begin{equation}\label{B_sigma=}
B_1=\left(
       \begin{array}{cc}
         \sqrt{1+\rho^2}\,\sin2\phi & \rho\,\cos2\phi \\[0.2cm]
         \rho & 0 \\
       \end{array}
     \right),~~~~~~
B_2=\left(
       \begin{array}{cc}
         \sqrt{1+\rho^2}\,\cos2\phi & -\rho\,\sin2\phi \\[0.2cm]
         0 & \sqrt{1+\rho^2}\\
         \end{array}
     \right).
\end{equation}

The integrability of the system \eqref{e'=} is provided by
$[A_1,\,A_2]=0$ and one gets the solution
\begin{equation}\label{e^A}
e=\exp (\xi^i A_i)\,\,C~,
\end{equation}
where $C$ is a constant $SO_\uparrow(2,2)$ matrix. The worldsheet
$Y(\s,\,\tau)$ can be read off from the second row of this solution.
In the following we explain how to calculate the exponential in
\eqref{e^A}, using the decomposition $SO_\uparrow(2,2)=SL(2,R) \times
SL(2,R)$.

The $SL(2,R)$ generators $t_\alpha$ $(\alpha=0,1,2)$ given by
\begin{equation}\label{t_alpha=}
t_0 =\left(
       \begin{array}{cc}
         0 & 1 \\
         -1 & 0 \\
       \end{array}
     \right),\quad
     t_1=\left(
            \begin{array}{cc}
              0 & 1 \\
              1 & 0 \\
            \end{array}
          \right),
\quad  t_2=\left(
            \begin{array}{cc}
              1 & 0 \\
              0 & -1 \\
            \end{array}
          \right),
\end{equation}
satisfy the relations
\begin{equation}\label{tt=}
t_\a t_\beta = \eta_{\a\beta}\, I+{\e_{\a\beta}}^\g \,t_\g~,
\end{equation}
where $\eta_{\a\beta}=\mbox{diag}(-1,1,1)$ is the metric tensor of 3d
Minkowski space, $I$ denotes the $2\times 2$ unit matrix and
$\e_{\a\beta\g}$ is the Levi-Civita tensor with $\e_{012}=1$.

Let us introduce $4\times 4$ matrices
\be\ba
L_0 &=\left(
       \begin{array}{cc}
         t_0 & 0 \\
         0 & t_0 \\
       \end{array}
     \right),
& L_1 &=\left(
            \begin{array}{cc}
              0 & t_1 \\
              t_1 & 0 \\
            \end{array}
          \right),
&L_2 &=\left(
            \begin{array}{cc}
              0 & t_2 \\
              t_2 & 0 \\
            \end{array}
          \right), \\[0.3cm]
R_0 &=\left(
       \begin{array}{cc}
         t_0 & 0 \\
         0 & -t_0 \\
       \end{array}
     \right),
&R_1 &=\left(
            \begin{array}{cc}
              0 & -t_0 \\
              t_0 & 0 \\
            \end{array}
          \right),
&R_2 &=\left(
            \begin{array}{cc}
              0 & I \\
              I & 0 \\
            \end{array}
          \right).
\ea\ee
They form a basis in $o(2,2)$ and satisfy the relations
(similar to \eqref{tt=})
\begin{equation}\label{Alg L,R}
L_\a L_\beta = \eta_{\a\beta}\,\hat{I}+ {\e_{\a\beta}}^\g\,
L_\g~,~~~~~~~~ R_\a R_\beta = \eta_{\a\beta}\,\hat{I}+{\e_{\a\beta}}^\g\, R_\g~,
\end{equation}
where $\hat I$ is the $4\times 4$ unit matrix. In addition, the matrices
$L_\a$ commute with $R_\b$, $\,\left[L_\a, R_\beta \right]=0$.
From \eqref{Alg L,R} follow the commutation relations of the
$sl(2)$ algebra
\begin{equation}\label{commutators}
\left[L_\a, L_\beta \right]=2{\e_{\a\beta}}^\g\,
L_\g~,~~~~~\left[R_\a,R_\beta \right]=2{\e_{\a\beta}}^\g\,R_\g~,
\end{equation}
and also simple exponentiation rules for arbitrary elements of $o(2,2)$.
In particular, the following relations holds
\be\ba\label{e^L=}
e^{\xi\,L_1}=\cosh\xi\,\,\hat I+\sinh\xi\,\,L_1~,\quad
e^{\theta\,L_0}=\cos\theta\,\,\hat I+\sin\theta\,\,L_0~,
\ea\ee
\begin{equation}\label{t-rotation}
e^{\frac{1}{2}\,\theta\,L_0}\,L_1\,e^{-\frac{1}{2}\,\theta\,L_0}=
\cos\theta\,L_1+\sin\theta\,L_2~,
\end{equation}
and similarly for the `right' part.

The matrices $A_i$ ($i=1,2$) in \eqref{matrices} are decomposed into
\begin{equation}\label{decompose}
A_i =l _{ij}\, L_j +r_{ij}\, R_j~,~~~~~~~~~~~~~ j=1,2~.
\end{equation}
One can read off the coefficients $l_{ij}$ and $r_{ij}$ from \eqref{B_sigma=},
and in matrix form one gets
\begin{equation}\label{l,r=}
l_{ij}=\left(\begin{array}{cc}
              \rho\cos^2\phi& \sqrt{1+\rho^2}\sin\phi\cos\phi \\[0.2cm]
              -\rho\sin\phi\cos\phi & -\sqrt{1+\rho^2}\sin^2\phi~ ~~
            \end{array}\right),~
r_{ij}=\left(\begin{array}{cc}
              \rho\sin^2\phi& \sqrt{1+\rho^2}\sin\phi\,\cos\phi \\[0.2cm]
              \rho\sin\phi\,\cos\phi & \sqrt{1+\rho^2}\cos^2\phi~ ~~~~
            \end{array}\right).
\end{equation}
Note that the rows of these two matrices are
proportional: $l_{2j}=-\tan\phi\,\,l_{1j}$, $\,r_{2j}=\cot\phi\,\,r_{1j}$.
This property leads to the vanishing of the commutator $[A_1,A_2]$,
and at the same time enables us to write $A_i$ in the form
\begin{equation}\label{A_i=}
A_i =e^{\frac{1}{2}\th_L L_0}\,e^{\frac{1}{2}\th_R R_0}
(l _i\, L_1 +r_i\, R_1)
e^{-\frac{1}{2}\th_L L_0}\,e^{-\frac{1}{2}\th_R R_0}~.
\end{equation}
Here we have used \eqref{t-rotation} and the fact that
$\theta_L$ and $\theta_R$ are independent of the rows.
These angles are defined by
\begin{equation}\label{theta_L=}
\tan\theta_L=\frac{\sqrt{1+\rho^2}}{\rho}\,\tan\phi~,~~~~~~
\cot\theta_R=\frac{\rho}{\sqrt{1+\rho^2}}\,\tan\phi~.
\end{equation}
Without loss of generality one can assume
$\phi\in\left(-\frac\pi 2,\frac\pi
2\right]$ and $\theta_L\in\left(-\frac\pi 2,\frac\pi
2\right]$, $\theta_R\in\left[-\pi,0\right)$.
Then, $\phi$ and $\theta_L$ have the same signs and
\begin{equation}\label{theta_L-theta_R}
-\pi <\theta_L+\theta_R <0~.
\end{equation}
Up to signs, the coefficients $l_i$ and $r_i$ in \eqref{A_i=}
correspond to the lengths of the rows $l_{ij}$ and $r_{ij}$,
respectively. They are given by
\be\ba\label{l_1=}
l_1&=\sqrt{\rho^2+\sin^2\phi}\,\,\cos\phi~,&
r_1&=-\sqrt{\rho^2+\cos^2\phi}\,\sin\phi~,\\
l_2&=-\sqrt{\rho^2+\sin^2\phi}\,\sin\phi~,&
r_2&=-\sqrt{\rho^2+\cos^2\phi}\,\cos\phi~.
\ea\ee
Here we have taken into account that $\sin\theta_L$ and $\sin\phi$
have the same signs,  $\cos\theta_R$ and $\sin\phi$
have opposite signs and, in addition,
$\cos\theta_L\geq 0,$ $~\sin\theta_R\leq 0$.

Due to \eqref{A_i=}, the exponential $e^{\xi^iA_i}$ can be written as
$e^B\,e^{\tilde{A}}\,e^{-B}$, where
\begin{equation}\label{B=,}
B=\frac 1 2\left(\th_L\, L_0+\th_R\, R_0\right)~,~~~~~~~~~
\tilde{A}=(l_i\,\xi^i)\,L_1+(r_i\,\xi^i)\,R_1~.
\end{equation}
The exponentials
$e^{\tilde{A}}$ and $e^B$ are obtained from \eqref{e^L=}.
In particular,
\begin{equation}\label{hat e=}
e^{\tilde{A}}= \left(
                        \begin{array}{cccc}
                          \cosh\eta & 0 & 0 & \sinh\eta \\
                          0 & \cosh\xi & \sinh\xi & 0 \\
                          0 & \sinh\xi & \cosh\xi & 0 \\
                          \sinh\eta & 0 & 0 & \cosh\eta \\
                        \end{array}
                      \right)~,
\end{equation}
where
\begin{equation}\label{eta,xi}
\xi=(l_i+r_i)\xi^i~,~~~~~~~\eta=(l_i-r_i)\xi^i~.
\end{equation}
The choice $C=e^B$ in \eqref{e^A} simplifies the form of the solution.
The corresponding  $Y(\sigma,\tau)$ is given as
a multiplication
of the matrix \eqref{hat e=} from the left side by the second row of
the matrix $e^B$. This row is $(\sin\theta,\, \cos\theta,\, 0,\,0)$,
with $\theta=-\frac1 2(\theta_L+\theta_R)$, and one finds
\begin{equation}\label{app Y=}
Y^n=(\sin\theta\cosh\eta,\,\, \cos\theta\cosh\xi,\,\,
\cos\theta\sinh\xi,\,\, \sin\theta\sinh\eta)~.
\end{equation}
The normal vector is obtained in a similar way as
\begin{equation}\label{N=}
N^n=(\cos\theta\cosh\eta,\,\, -\sin\theta\cosh\xi,\,\,
-\sin\theta\sinh\xi,\,\, \cos\theta\sinh\eta)~.
\end{equation}
In the main text we use the  vector $Y$ with covariant indices.
Therefore, the solution \eqref{AdS sol} is related to \eqref{app Y=}
by the isometry transformation with the matrix
$\mbox{diag}(-1,-1,1,1).$

From \eqref{theta_L-theta_R} follows that $\theta\in\left(0,\frac\pi 2\right)$.
The calculation of $\cot2\theta$ with the help of \eqref{theta_L=}
yields
\begin{equation}\label{cot2th}
\cot2\theta=\frac{\sin2\phi}{2\rho\sqrt{1+\rho^2}}~,
\end{equation}
and by \eqref{R=} it takes the form \eqref{theta=}.

Eq. \eqref{eta,xi} defines $\xi$ and $\eta$ as a
linear transformation \eqref{eta,xi=} with
\begin{equation}\label{A,B}
A=l_1+r_1~,~~~~B=l_2+r_2~,~~~~~C=l_1-r_1~,~~~D=l_2-r_2~.
\end{equation}
Calculating the squares of these coefficients one can easily eliminate
the $\phi$ variable through \eqref{cot2th} and
get \eqref{A,B=}.
On the other hand, the signs of the coefficients \eqref{A,B}
depend on values of $\phi$ by \eqref{l_1=}.
For example, if $\phi\in\left(-\frac\pi 2,
-\frac\pi 4\right)$, one has $A>0$, $B>0$, $C<0$, $D>0$,
which corresponds to the case \eqref{A,B=}.
Its continuation to other intervals $\phi\in\left(-\frac\pi 4,
\frac\pi 4\right)$ and $\phi\in\left(\frac\pi 4,
\frac\pi 2\right)$  is straightforward.

\setcounter{equation}{0}
\def\theequation{B.\arabic{equation}}
\section{Integration of the linear system in $\rr^4$}
\label{AppSphere}

In the spherical case we use the decomposition $SO(4)=SO(3)\times
SO(3)$. The matrices
\be\ba\label{O(4)=}
\tilde{L}_1 &=\left(
            \begin{array}{cc}
              0 & t_1 \\
              -t_1 & 0 \\
            \end{array}
          \right),&
\tilde{L}_2 &=\left(
            \begin{array}{cc}
              0 & t_2 \\
              -t_2 & 0 \\
            \end{array}
          \right),&
\tilde{L}_3 &=\left(
       \begin{array}{cc}
         t_0 & 0 \\
         0 & t_0 \\
       \end{array}
     \right),\\[0.3cm]
\tilde{R}_1 &=\left(
            \begin{array}{cc}
              0 & t_0 \\
              t_0 & 0 \\
            \end{array}
          \right),&
\tilde{R}_2 &=\left(
            \begin{array}{cc}
              0 & -I \\
              I & 0 \\
            \end{array}
          \right),&
\tilde{R}_3 &=\left(
       \begin{array}{cc}
         t_0 & 0 \\
         0 & -t_0 \\
       \end{array}
     \right),
\ea\ee
are antisymmetric and satisfy the relations
\begin{equation}\label{o(4) Alg L,R}
\tilde L_a \tilde L_b =-\delta_{ab}\,\hat I+
\e_{abc}, \tilde L_\g~,~~~~~~~~ \tilde R_a\,\tilde
R_b =-\delta_{ab}\,\hat I+\e_{abc} \tilde R_c,
\end{equation}
where the indices $a,b,c$ range from 1 to 3.
In addition, the commutators $[\tilde L_a,\,\tilde R_b]$
vanish.

Then, using the same scheme as in $AdS_3$ one finds
the solution \eqref{sph sol} up to $SO(4)$ transformations.

\setcounter{equation}{0}
\def\theequation{C.\arabic{equation}}
\section{Calculation of the regularized area}\label{AppArea}
Here we analyze the integral \eqref{I(r_c)}.
The domain of integration is bounded by the line \eqref{contour}
and due to the symmetry of the contour we have $I(r_c)=2 I$,
where $I$ is the area bounded in the
first quarter of the ($\xi,\,\eta)$ plane
and it is given by
\begin{equation}\label{I}
I=\int_0^{\xi_c} \mbox{d}\xi\,\,\eta(\xi)~.
\end{equation}
Introducing the integration variable $u=\cosh\xi$, we find
\begin{equation}\label{I_epsilon}
I=\int_1^{\frac{1-\epsilon}{\epsilon'}}\frac{\mbox{d}u}
{\sqrt{u^2-1}}\,\,\log\left[\frac{1-\epsilon'
u}{\epsilon}+\sqrt{\left(\frac{1-\epsilon' u}{\epsilon}\right)^2-1}\,\right]~.
\end{equation}
We split this integral into two terms $I=I_1+I_2$
with
\begin{equation}\label{I_1}
I_1=-\log\epsilon \int_1^{\frac{1-\epsilon}{\epsilon'}}\frac{\mbox{d}u}
{\sqrt{u^2-1}}~,
\end{equation}
\begin{equation}\label{I_2}
I_2=\int_1^{\frac{1-\epsilon}{\epsilon'}}\frac{\mbox{d}u}
{\sqrt{u^2-1}}\,\,\log\left[1-\epsilon'
u+\sqrt{\left(1-\epsilon' u\right)^2-\epsilon^2}\,\right]~.
\end{equation}
The integration in \eqref{I_1} yields
\begin{equation}\label{I_1=}
I_1=\log\epsilon\,\log\epsilon'
-\log\epsilon\,\log\left[1-\epsilon+\sqrt{(1-\epsilon)^2-\epsilon'^2}\right]~,
\end{equation}
and neglecting the vanishing terms at $r_c\rightarrow 0$
($\epsilon\rightarrow0$, $\epsilon'\rightarrow0$), we find
\begin{equation}\label{I_1appr}
I_1\simeq \log\epsilon\left(\log\epsilon'-\log 2\right)~.
\end{equation}
A similar approximation in \eqref{I_2} provides
\begin{equation}\label{I_2=}
I_2=\int_1^{\frac{1}{\epsilon'}}\frac{\mbox{d}u}
{\sqrt{u^2-1}}\,\,\log \left[2\left(1-\epsilon'u\right)\right]~,
\end{equation}
and it splits into two parts $I_2=I_3+I_4$, with
\begin{equation}\label{I_3}
I_3=\log2\int_1^{\frac{1}{\epsilon'}}\frac{\mbox{d}u}
{\sqrt{u^2-1}}~,~~~~~~
I_4=\int_1^{\frac{1}{\epsilon'}}\frac{\mbox{d}u}
{\sqrt{u^2-1}}\,\,\log \left(1-\epsilon'u\right)~.
\end{equation}
In a same way as in \eqref{I_1},
\begin{equation}\label{I_3=}
I_3\simeq \log2\left(\log2-\log\epsilon'\right)~.
\end{equation}
Changing the integration variable in $I_4$ by $\epsilon'u=x$, we get
\begin{equation}\label{I_4=}
I_4=\int_{\epsilon'}^1 \frac{\mbox{d}x}{\sqrt{x^2-\epsilon'^2}}
\,\,\log (1-x) \simeq \int_0^1 \frac{\mbox{d}x}{x} \log(1-x)=-\frac{\pi^2}{6}.
\end{equation}
As a result we obtain
\begin{equation}\label{I_epsilon=}
I(r_c)\simeq 2(\log2-\log\epsilon)(\log2-\log\epsilon')-\frac{\pi^2}{3}\,.
\end{equation}

Inserting $\epsilon$ and $\epsilon'$ from \eqref{epsilon=}, we arrive
at \eqref{I(r_c)=}.

\end{document}